\documentclass{jpp}
\usepackage{graphicx}
\usepackage{amsmath}
\usepackage{amsfonts}
\usepackage{amssymb}
\usepackage{epstopdf, epsfig}
\usepackage{xcolor}
\usepackage{bigints}
\usepackage{verbatim}
\usepackage{bm}

\definecolor{GMcolor}{rgb}{0.1,0.1,1}
\definecolor{JBcolor}{rgb}{0.5,0.1,1}

\usepackage[utf8]{inputenc}
\usepackage[T1]{fontenc}
\usepackage{amsmath}

\shorttitle{Guidelines for authors}
\shortauthor{G. Miloshevich, J. W. Burby}

\title{Hamiltonian reduction of Vlasov-Maxwell to a dark slow manifold}

\author{George Miloshevich\aff{1}
  \corresp{\email{george.miloshevich@ens-lyon.fr}}
 \and Joshua W. Burby\aff{2}}

\affiliation{\aff{1}\'Ecole Normale Sup\'erieure de Lyon \\
	Laboratoire de Physique, 46, allée d'Italie, F-69364, Lyon cedex 07, France\\
	\aff{2}Los Alamos National Laborat ory, Los Alamos, New Mexico 87545, USA}

\newcommand{\pr}{\partial}
\newcommand{\f}[2]{\frac{#1}{#2}}
\newcommand{\lf}{\left}
\newcommand{\ri}{\right}
\newcommand{\dd}[2]{\frac{d #1}{d #2}}
\newcommand{\pp}[2]{\frac{\pr #1}{\pr #2}}

\newcommand{\delf}[2]{\frac{\delta #1\hfill}{\delta #2\hfill}}

\newcommand{\calF}{\mathcal{F}}
\newcommand{\calG}{\mathcal{G}}

\newcommand{\calJ}{\mathcal{J}}

\newcommand{\calM}{\mathcal{M}}

\newcommand{\calP}{\mathcal{P}}

\begin{document}

\maketitle

\begin{abstract}
    We show that nonrelativsitic scaling of the collisionless Vlasov-Maxwell system implies the existence of a formal invariant slow manifold in the infinite-dimensional Vlasov-Maxwell phase space. Vlasov-Maxwell dynamics restricted to the slow manifold recovers the Vlasov-Poisson and Vlasov-Darwin models as low-order approximations, and provides higher-order corrections to the Vlasov-Darwin model more generally. The slow manifold may be interpreted to all orders in perturbation theory as a collection of formal Vlasov-Maxwell solutions that do not excite light waves, and are therefore ``dark." We provide a heuristic lower bound for the time interval over which Vlasov-Maxwell solutions initialized optimally-near the slow manifold remain dark. We also show how dynamics on the slow manifold naturally inherit a Hamiltonian structure from the underlying system. After expressing this structure in a simple form, we use it to identify a manifestly Hamiltonian correction to the Vlasov-Darwin model. The derivation of higher-order terms is reduced to computing the corrections of the system Hamiltonian restricted to the slow manifold.
 
\end{abstract}


 \section{Introduction}
The Vlasov-Maxwell, Vlasov-Poisson, and Vlasov-Darwin models are three of the most well-known kinetic descriptions of fully-ionized plasmas. In CGS units, the Vlasov-Poisson system can be "derived" from the Vlasov-Maxwell system by sending the {normalized} speed of light $c$ to infinity. The Vlasov-Darwin model can then be understood as the next-to-leading order correction to Vlasov-Poisson when expanding Vlasov-Maxwell in powers of $1/c$. This picture of the interrelationships between these three models, while sufficient for many purposes in plasma modeling, begs a number of subtle questions. In particular,
\begin{itemize}
    \item[(1)] In what sense does a solution of the Vlasov-Poisson system or the Vlasov-Darwin system approximate a solution of the Vlasov-Maxwell system? How accurate is the approximation, and for how long is the approximation valid?
    \item[(2)] The Vlasov-Darwin model improves the Vlasov-Poisson model; can one improve Vlasov-Darwin without invoking the full complexity of Vlasov-Maxwell?
    \item[(3)] In the absence of collisions, each of these models has its own Hamiltonian structure. How can the Vlasov-Poisson and Vlasov-Darwin structures be deduced systematically from the Vlasov-Maxwell structure? If there are higher-order corrections to Vlasov-Darwin, what can be said of their Hamiltonian structures?
\end{itemize}

Question (1) has been studied intensively by a number of authors in the applied mathematics community, including \cite{Asano_1986}, \cite{Degond_1986}, and \cite{Schaeffer_1986}. Each of these authors independently showed that solutions of the Vlasov-Maxwell system with ``well-prepared" initial conditions remain within $1/c$ of corresponding Vlasov-Poisson solutions on bounded, $c$-independent time intervals. Here, ``well-prepared" means that the initial magnetic field vanishes and the initial electric field is irrotational. Similar results are given in \cite{Degond92} for the Vlasov-Darwin system, where it is shown that Vlasov-Darwin solutions provide second-order approximations to the magnetic field and third order approximations to the electric field predicted by Vlasov-Maxwell. Again, the approximations apply on bounded, $c$-independent time intervals for well-prepared initial data. A shortcoming of these analyses is their lack of a phase-space-geometric interpretation. In particular, the geometric significance of ``well-prepared" initial data is unclear.

Question (2) is less studied than (1), but plays a prominent role in \cite{HanKwan2018td}, where linearizations of high-order corrections to Vlasov-Darwin were used to establish long-term nonlinear stability of Penrose-stable Vlasov-Maxwell equilibria. This analysis clearly shows that better approximations follow from increasingly well-prepared initial data. Again, the phase-space-geometric picture of these results is lacking. Question (3) has also received fairly little attention in the literature. The most notable result in this direction comes from \cite{Chandre_2013}, who obtains the Poisson bracket for the Vlasov-Poisson system by applying Dirac constraint theory to the Vlasov-Maxwell bracket. It is unclear if this construction extends to treat the Vlasov-Darwin system, or to corrections thereof.

The purpose of this Article is to develop a phase-space-geometric picture of the answers to each of these questions. Our fundamental observation will be that the electrostatic non-relativistic scaling of the Vlasov-Maxwell system \eqref{scaling_formulas} gives rise to a formal invariant manifold in the Vlasov-Maxwell phase space. By calculating the asymptotic expansion of this manifold, we will demonstrate that Vlasov-Maxwell dynamics on the manifold coincides with Vlasov-Poisson dynamics to leading order, and Vlasov-Darwin dynamics at next-to-leading order. More generally, higher-order truncations of dynamics on the manifold lead to nonrelativistic plasma models that improve on Vlasov-Darwin and share a phase space with the Vlasov-Poisson system. That is, they comprise closed evolution laws on the space of distribution functions, rather than on the composite space of distribution functions and electromagnetic fields in which Vlasov-Maxwell dynamics evolves in general.

These observations immediately shed light on questions (1) and (2) above. Solutions of the Vlasov-Poisson and Vlasov-Darwin models should be understood as low-order approximations of Vlasov-Maxwell dynamics on its (formal) invariant manifold. Better approximations can be found systematically by deriving the higher-order on-manifold corrections to Vlasov-Maxwell. In particular, better approximations require choosing initial conditions that lie on the formal invariant manifold with greater accuracy, which explains and generalizes the role played by ``well-prepared initial data" in the work referenced above. The timescale over which such approximations are valid can be no longer than the normal stability timescale for the formal invariant manifold.

Physically, the formal invariant manifold represents nontrivial plasma motions that are free of light waves generated by collective effects. Thus, phase space points on the formal invariant manifold represent formal solutions of the Vlasov-Maxwell system that are \emph{dark} to all orders in $1/c$. Here we are careful to distinguish between collective light waves and light waves more generally since plasmas exhibit emission processes \cite{bekefi66} that are not captured by the Vlasov-Maxwell model. We also emphasize that here we refer specifically to light waves, i.e. waves with the dispersion relation $\omega^2 = \omega_p^2 + k^2\,c^2$ in the small-amplitude regime, and not more general 
{fast modes} such as compressional Alfv\'en waves. Finally, we stress that ``formal solutions" differ from ``solutions," since the formal invariant manifold is not a true invariant object in the Vlasov-Maxwell phase space -- truncations of its asymptotic expansion provide approximate invariant manifolds. Therefore, in general, we should only expect solutions of the Vlasov-Maxwell system initialized near the formal invariant set to remain dark for a limited interval of time. In Section \ref{dark_stability_sec}, we will give a heuristic argument based on rigorous analysis from \cite{kristiansen16} that this time interval is likely at least $O(c)$ as $c\rightarrow \infty$ for Vlasov-Maxwell solutions initialized on an optimal truncation of the formal invariant manifold.

Mathematically, the formal invariant manifold underlying Vlasov-Poisson, Vlasov-Darwin, and higher-order corrections thereof is an example of a slow manifold. The general theory of slow manifolds is reviewed in \cite{mackay04} for a mathematical audience and in \cite{burby_review_2020} for an audience of plasma physicists. Since a slow manifold in a Hamiltonian system necessarily inherits a Hamiltonian structure, it follows immediately that the dark slow manifold in the Vlasov-Maxwell system has a natural Hamiltonian structure. We will calculate this induced Hamiltonian structure and show that it reproduces the known Hamiltonian structures underlying the Vlasov-Poisson and Vlasov-Darwin systems. Moreover, we will introduce a formal near-identity change of dependent variables on the slow manifold that makes the slow manifold Poisson bracket agree with the Vlasov-Poisson bracket to all orders in $1/c$. The same procedure, which we call symplectic rectification, was used by \cite{burby2017magnetohydrodynamic} to obtain a closed-form expression for the bracket on the all-orders extended MHD slow manifold inside of the two-fluid system. In terms of these rectified variables, we will then formulate a manifestly-Hamiltonian correction to the Vlasov-Darwin model. See Eq.\,\eqref{dark_ham} for the post-Darwin Hamiltonian, Eqs.\,\eqref{darkbracket} for the Poisson bracket, and Eq.\,\eqref{dark_hamiltonian_vlasov}-\eqref{longitudinal_vector_potential} for the definitions of the rectified variables. To the best of our knowledge, this is the first Hamiltonian correction to Vlasov-Darwin that appears in the literature. Deducing higher-order Hamiltonian corrections is reduced to the task of computing higher-order terms in the system Hamiltonian restricted to the slow manifold.  Altogether, these results provide a satisfying resolution of question (3) above.

The outline of this Article is as follows. First we briefly review
slow manifold reduction in Section \ref{The mathematics of slow manifold reduction}. We observe that the theory of slow manifold reduction applies to the Vlasov-Maxwell system in Section \ref{The Vlasov-Maxwell equations as a fast-slow system}, and then exploit this observation in Section \ref{The dark slow manifold} in order to formally demonstrate the existence of dark plasmas and uncover their governing dynamical laws. The rest of the article, i.e. Section \ref{Dark plasma dynamics as a Hamiltonian system} will be devoted to a rather detailed analysis of the basic structural properties of dark plasma dynamics. We will show: (a) the Hamiltonian structure underlying the Vlasov-Maxwell equations is inherited by dark plasma dynamics (Section \ref{Why is dark plasma dynamics Hamiltonian?}); (b) a formally-exact expression for the dark plasma Poisson bracket may be obtained through the application of a sequence of near-identity changes of dependent variables we call symplectic rectification (Section \ref{Derivation of the rectifying transformation}); (c) the rectifying transformation may be calculated efficiently using infinite-dimensional Lie transforms; and (d) explicit expressions for the dark plasma Poisson bracket and Hamiltonian functional (Section \ref{Calculation of the dark Poisson bracket and post-Darwin Hamiltonian}). 

\section{The Vlasov-Maxwell equations as a fast-slow system}
\subsection{Slow manifold reduction\label{The mathematics of slow manifold reduction}}
We will apply a dimension-reduction technique for dynamical systems with a time scale separation known as slow manifold reduction. An overview of this tool, tailored for an audience of plasma physicists, can be found in \citet{burby2020}, while a more mathematical review is given by \citet{mackay04}. { Readers can also consult~\citet{VANKAMPEN198569} for additional applications and context. The reduction to slow manifold has been used historically by~\citet{lorenz86} in the hopes of understanding how fast internal gravity dynamics can be decoupled from the slow Rossby waves consistent with meteorological observations to explain why the latter dominate. This was immediately followed by proving the non-existence of the aforementioned manifold in~\citet{Lorenz87} and the clarification~\citet{lorenz92}.  We note that when slow manifold existence and stability cannot be established rigorously for a given mathematical model one may consider a ``backwards'' theory alternative, e.g.~\citet{roberts15}, which states that there exists a system close to the original one with an exact slow manifold. } 

Let the system be describable by a set of ``slow'' variables $x$ and ``fast'' variables $y$, and suppose that the dynamics of the pair $(x,y)$ are prescribed by the system of (possibly infinite-dimensional) ordinary differential equations
\begin{eqnarray}\label{slowfastsystem}
\dot x = g_\epsilon(x,y)\nonumber\\
\epsilon\dot y = f_\epsilon(x,y),
\end{eqnarray}
where $0 < \epsilon \ll 1$ is a small parameter. When $\epsilon = 0$, we see that the second differential equation 
degenerates to an algebraic equation that imposes the constraint $y = y_{0}^\star(x)$, where $y_{0}^\star$ is defined implicitly by
\begin{equation}
f_{0}(x, y_{0}^\star(x)) = 0.
\end{equation}
It was shown by~\citet{Fenichel1979} in finite dimensions that if the manifold
\begin{equation}
\calM = \{ (x,y): y = y_{0}^\star \}
\end{equation}
is normally hyperbolic (all eigenvalues of $D_y f_\epsilon (x,y^\star(x))$ have non-zero real parts) then, under some regularity assumptions, the dynamics of $(x,y)$ with $1\gg\epsilon >0$ contain an invariant manifold that is close to $\calM$ and that converges to $\calM$ as $\epsilon\rightarrow 0$. 

When $\calM$ is not normally hyperbolic, the invariant manifold need not exist. However, \emph{almost} invariant manifolds can still be constructed by introducing the formal power series
\begin{equation}\label{yepxansion}
y^\star(x) = y^\star_{0} + \epsilon y^\star_{1} + \epsilon^2 y^\star_{2} + \dots,
\end{equation}
and demanding that the ``graph" of $y^\star$ is invariant order-by-order in $\epsilon$. For non-negative integer $n$, the manifold
\begin{equation}
\calM_\epsilon^{(n)} = \{(x,y)\, |\, y = (y_0^\star + \dots + \epsilon^n y_n^*)(x)\}
\end{equation}
is almost invariant in the sense that the normal component of the vector field $(\dot{x},\dot{y})$ is small along $\calM_\epsilon^{(n)}$. Because the projection of $(\dot{x},\dot{y})$ along $\calM_\epsilon^{(n)}$ is $O(1)$, $\calM_\epsilon^{(n)}$ is known as a \emph{slow manifold of order} $n$, while the formal power series $y^\star(x)$ is known as the \emph{formal slow manifold} or the \emph{slaving function}.

The various terms of the formal slow manifold are obtained from the appropriate ordering of the original equations of motion~\eqref{slowfastsystem}. One introduces Fr\'{e}chet derivatives  (for background see e.g. \cite{courant53,Lang_1995}), {which are commonly computed using the directional derivative formula}
\begin{equation}
DF(\psi)\lf[\delta\psi \ri] := \dd{}{\epsilon}\Big|_0 F(\psi + \epsilon\,\delta \psi),
\end{equation}
the algorithm for obtaining these terms amounts to solving the partial differential equation (functional partial differential equation in infinite dimensions) 	
\begin{equation}
\epsilon D_x y^\star_\epsilon(x) \lf[g_\epsilon(x,y^\star_\epsilon(x)) \ri] = f_\epsilon(x, y_\epsilon^\star(x)),
\end{equation}
order-by-order in $\epsilon$.
For instance, the first order terms are
\begin{equation}
D_x y^\star_{0}(x) \lf[g_{0}(x,y^\star_{0}(x)) \ri] = f_{1}(x, y_{0}^\star(x)) + D_y f_{0}(x,y^\star_{0}(x)) \lf[y_{1}^\star(x) \ri]
\end{equation}
Likewise, the second order terms, albeit somewhat cumbersome, are written 
\begin{equation}\begin{aligned}
&\varepsilon^{2} \Rightarrow D y_{1}^{*}(x)\left[g_{0}\left(x, y_{0}^{*}(x)\right)\right]+D y_{0}^{*}(x)\left[g_{1}\left(x, y_{0}^*\right)(x)\right]+\\
&+D y_{0}^{*}(x)\left[D_y g_{0}\left(x, y_{0}^{*}(x)\right)\left[y_{1}^{*}(x)\right]\right]=f_{2}\left(x, y_{0}^{*}(x)\right)+ \\
&\begin{array}{l}
D_{y} f_{1}\left(x, y_{0}^{*}(x)\right)\left[y_{1}^{*}(x)\right]+\dfrac{1}{2} D_{y}^{2} f_{0}\left(x, y_{0}^{*}(x)\right)\left[y_{1}^{*}(x), y_{1}^{*}(x)\right]+ \\
+D_{y} f_{0}\left(x, y_{0}^{*}(x)\left[y_{2}^{*}(x)\right]\right.
\end{array}
\end{aligned}\end{equation}

It is natural to wonder about the normal stability of the slow manifolds $\calM_\epsilon^{(n)}$. In other words, if a trajectory $(x(t),y(t))$ begins near $\calM_\epsilon^{(n)}$ then how long does it remain near $\calM_\epsilon^{(n)}$? When $\calM$ is attractive, one expects stability over arbitrarily-large time intervals. When $\calM$ is normally elliptic (imaginary eigenvalues), as will be the case for the slow manifold studied in this Article, dynamics near a slow manifold are neutrally-stable to leading-order in perturbation theory, and therefore exhibit normal stability on at least an $O(1)$ timescale (Note that the shortest timescale contained in the dynamical system \eqref{slowfastsystem} is $O(\epsilon)$). However, higher-order effects can lead to resonances that destabilize the slow manifold on $O(1/\epsilon)$-timescales. For a large class of Hamiltonian fast-slow systems with finitely-many slow variables and elliptic slow manifolds, \citet{kristiansen16} showed that there is an optimal truncation of the slow manifold that remains normally stable on the $O(1/\epsilon)$ timescale. Establishing normal stability in the elliptic case over even larger time intervals may sometimes be achieved by identifying an adiabatic invariant whose critical manifold coincides with the formal slow manifold, as in \citet{Cotter2004}. Long-term slow manifold stability is also consistent with the long-term equilibrium stability results obtained for the Vlasov-Maxwell system with non-relativistic scaling in \cite{HanKwan2018td}. In the spirit of these previous studies, we conjecture that the slow manifold identified in this Article exhibits normal stability on an $O(1/\epsilon)$ timescale, at least when optimally truncated. In Section \ref{dark_stability_sec} we outline how the optimal truncation strategy developed in \cite{kristiansen16} applies formally to the Vlasov-Maxwell system, and therefore provide additional supporting evidence for our conjecture. In future work, we plan to investigate the possibility of our slow manifold arising as the zero level set of a wave action adiabatic invariant, which would provide further insight into normal stability.  


\subsection{The Vlasov-Maxwell equations as a fast-slow system\label{The Vlasov-Maxwell equations as a fast-slow system}}

Various plasma models, including magnetohydrodynamics (MHD) and Hall MHD, rely on the assumption that one can ignore certain {large-scale} high-frequency modes, such as light waves. This drastically reduces the multi-scale difficulty of modeling a plasma. However, the approximation is often made in ad hoc fashion. Here we work with the collisionless Vlasov-Maxwell system of equations and systematically perform an asymptotic expansion such that formally the small parameter corresponds to the  inverse {normalized} speed of light $c^{-1}$. In 
{particular}, in order to form a dimensionless small parameter one can consider a ratio between a typical speed of a particle and the speed of light. This expansion procedure bears a resemblance to the well-known Chapman-Enskog method, but retains the connection with the (infinite-dimensional) phase space geometry underlying the closure. This is achieved by computing the slaving function $y^\star(x)$ instead of power-series expansions of solutions of the Vlasov-Maxwell equations. This more geometric perspective on the closure problem will be instrumental in uncovering the closure's Hamiltonian structure.
 
We assume that each plasma species $\sigma$ (typically two species are considered, but any number of species may be included without difficulty) obeys a collisionless relativistic Vlasov equation  
\begin{equation}\label{si:maxwellvlasov}
\pp{f_\sigma}{t} + {\bf v}\cdot \pp{f_\sigma}{\bf r}  + e_\sigma\lf( {\bf E}  + {\bf v}\times {\bf B} \ri) \cdot \pp{f_\sigma}{{\bf p}} = 0, 
\end{equation}
coupled, as usual, with a self-consistent electromagnetic field defined on the triply-periodic spatial domain $\mathbb{Q} = \mathbb{S}^1\times \mathbb{S}^1\times \mathbb{S}^1$. (Our slow manifold analysis is straightforward to adapt to other domain topologies and boundary conditions. However, our analysis of the Hamiltonian structure on the slow manifold requires either an unbounded or periodic domain. The reason for this limitation is that the Hamiltonian structure underlying the Vlasov-Maxwell system is only understood in unbounded or periodic domains; identifying the Hamiltonian structure in a bounded domain with even simple boundary conditions like specular reflection remains an open research problem.)
The field evolves according to Maxwell's equations in SI units
\begin{equation}\label{displacementcurrent}
    -\frac{1}{c^{2}} \frac{\partial {\bf E}}{\partial t}+{\nabla} \times {\bf B}=\mu_{0} {\bf J} - \mu_0\mathbf{J}_H,
\end{equation}
\begin{equation}\label{faradaylaw}
    \frac{\partial {\bf B}}{\partial t}+\nabla \times {\bf E}=0,
\end{equation}
\begin{equation}\label{gausslaw}
    {\nabla} \cdot {\bf E}=\frac{\rho - \rho_H}{\varepsilon_{0}}\quad\text{and}\quad \nabla \cdot {\bf B} = 0,
\end{equation}
where $\mu_{0}\varepsilon_0=1/{c^{2}}$ {with $c$ a dimensionful speed of light}, $\rho_H = \int_{\mathbb{Q}} \rho\,d^3r / \int_{\mathbb{Q}}\,d^3r $, and $\mathbf{J}_H = \int_{\mathbb{Q}}\,\mathbf{J}\,d^3r/\int_{\mathbb{Q}}\,d^3r$ denotes the harmonic component of the current density (for more details see Appendix~\ref{hhdecompose}). {Note that the integral of the usual Gauss' Law $\epsilon_0\,\nabla\cdot \mathbf{E} = \rho $ over the periodic domain $Q = (\mathbb{S}^1)^3$ imposes the restriction $\int_Q\rho\,d^3x = 0$, i.e. that the total charge vanishes. Therefore the $\rho_H$ in \eqref{gausslaw} may be interpreted as the constant neutralizing background necessary to enforce this consraint when the plasma itself has a net charge.}
In order to find the small parameter, we pass to new dimensionless variables via a transformation that identifies typical length $L_0$ and time $t_0$ scales
\begin{gather}
    t_0 = \sqrt{\frac{4\pi\,\varepsilon_0m_0}{e_0\,\rho_0}}\nonumber\\
    m_\sigma = m_0 m_\sigma^\prime,\quad e_\sigma = e_0\,e_\sigma^\prime,\quad t =  t_0 t^\prime,\nonumber\\
    f_\sigma(\mathbf{r},\mathbf{p}) = \frac{\rho_0}{e_0}\left(\frac{t_0}{m_0\,L_0}\right)^3\,f_\sigma^\prime\left(\frac{\mathbf{r}}{L_0},\frac{t_0\mathbf{p}}{m_0\,L_0}\right) ,\nonumber\\
     \mathbf{E}(\mathbf{r}) = \frac{L_0\,\rho_0}{4\pi\varepsilon_0}\mathbf{E}^\prime\left( \frac{\mathbf{r}}{L_0} \right),\quad \mathbf{B}(\mathbf{r}) = \frac{L_0\,\rho_0}{4\pi\varepsilon_0\,c}\,\mathbf{B}^\prime\left(\frac{\mathbf{r}}{L_0}\right)\label{scaling_formulas}
\end{gather}

After the transformation is applied, primes are suppressed and a small parameter is identified as $\epsilon := L_0/(c t_0)$. In what follows, we will use the symbol $1/c$ as a convenient placeholder for $\epsilon$.  The reader may observe that in this way the equations appear to be written in CGS units, which are better suited for the particular non-relativistic limit we study. After application of these steps the Vlasov equation~\eqref{si:maxwellvlasov} is re-written
\begin{equation}\label{maxwellvlasov}
\dot{f}_\sigma + {\bf v}\cdot \nabla f_\sigma + e_\sigma\lf( {\bf E}  + \f{1}{c}{\bf v}\times {\bf B} \ri) \cdot \pp{f_\sigma}{{\bf p}} = 0, 
\end{equation}
where the electric field ${\bf E} \in \Omega^{1}_T\oplus\Omega^{1}_L $ is a vector field on $\mathbb{Q}$ and can be decomposed into transverse and longitudinal components ${\bf E} = {\bf E}_T + {\bf E}_L =: \Pi_L {\bf E} + \Pi_T {\bf E}$,  such that $\nabla\cdot {\bf E}_T = 0$ and $\nabla\times {\bf E}_L = 0$ {See Appendix~\ref{hhdecompose} for a review of Helmholtz-Hodge decomposition on compact manifolds without boundaries}. (The most general decomposition of a smooth vector field on the triply-periodic domain $\mathbb{Q}$ must allow also for a harmonic component $\mathbf{E}_H$ such that $\nabla\cdot\mathbf{E}_H = 0$ and $\nabla\times\mathbf{E}_H = 0$. However, this harmonic component may be dropped self-consistently provided the harmonic component of the current density is subtracted from the Amp\'ere-Maxwell equation, as in Eq.\,\eqref{displacementcurrent}.) Note that the projection operators may be expressed as
\begin{eqnarray}
\Pi_T = 1 - \Pi_L - \Pi_H  = -\Delta^{-1}\nabla\times\nabla\times \\ \Pi_L = \nabla\Delta^{-1}\nabla\cdot
\end{eqnarray}
For details see~\eqref{PiLEquals} and~\eqref{PiTEquals}.

Single-particle phase-space distribution functions $f_\sigma \in L^1(T^\star \mathbb{Q})\cap C^\infty(T^*\mathbb{Q})$ are integrable and smooth on $T^\star \mathbb{Q}= \mathbb{Q}\times \mathbb{R}^3$. Moreover, ${\bf p} = m_\sigma\gamma_\sigma {\bf v}$, where
\begin{equation}\label{gammafactor}
\gamma_\sigma = \sqrt{1+\f{p^2}{m^2_\sigma c^2}}.
\end{equation}
The magnetic field ${\bf B}\in \Omega_T^2$ is purely transverse $\nabla\cdot {\bf B} = 0$. Likewise in the new units we have~\eqref{gausslaw} a Gauss law constraint
\begin{equation}\label{gausslawconstraint}
    {\nabla} \cdot {\bf E}_L=4\pi \sum_\sigma e_\sigma\int d^3 p\, f_\sigma - 4\pi\rho_H.
\end{equation}

For the transverse and longitudinal components of the electric field we have the evolution equations
\begin{equation}\label{Emaxwell2}
\f{1}{c} \pp{{\bf E}_L}{t} = -\f{4\pi}{c} {\bf J}_L, 
\end{equation}
\begin{equation}\label{Emaxwell}
\f{1}{c} \pp{{\bf E}_T}{t} = \nabla\times{\bf B}-\f{4\pi}{c} {\bf J}_T,
\end{equation}
where
\begin{equation}
{\bf J}  := \sum_\sigma e_\sigma\int d^3 p \, {\bf v} f_\sigma
\end{equation} 

{Using the gauge where the electrostatic potential vanishes,} the relationship between ${\bf E}_L$ and the longitudinal part of the vector potential ${\bf A}_L$ is
\begin{equation}
    {\bf E}_{L}=-\frac{1}{c} \frac{\partial {\bf A}_{L}}{\partial t}.
\end{equation}
Finally, the system is closed by Faraday law
\begin{equation}\label{BFaraday}
\f{1}{c} \pp{{\bf B}_T}{t} = - \nabla\times {\bf E}_T.
\end{equation}

It has been proved by~\citet{Degond92} that the Darwin
model in three dimensional cases approximates Maxwell’s equations for appropriate initial data up to the second order of the dimensionless parameter $v/c$ for magnetic field $B$ and to the third order for electric field $E$, where $v$
is the characteristic velocity. (The somewhat mysterious notion of "appropriate" initial data is laid bare in the context of slow manifold reduction theory; well-prepared initial data are merely initial conditions chosen to lie on the slow manifold.) In what follows we will re-derive this result and go an order beyond Darwin to establish the next approximation.

\subsection{The dark slow manifold: Piezoelectric correction\label{The dark slow manifold}}

The candidates for the slow variables in our case are $X = (L^{1}(T^\star \mathbb{Q})\cap C^\infty(T^*\mathbb{Q}))\times\Omega_L^1 \ni (f, {\bf E}_L)$  and, following the recipe \eqref{yepxansion}, we treat transverse fields as fast $Y = \Omega_T^1\times\Omega_L^2 \ni ({\bf E}_T,{\bf B})$ (which physically carry light waves.) Our slaving functions are therefore 
\begin{align}
{\bf E}^\star_T(f_\sigma,\mathbf{E}_L) &= 	{\bf E}^\star_{T0}(f_\sigma,\mathbf{E}_L) + \f{1}{c}{\bf E}^\star_{T1}(f_\sigma,\mathbf{E}_L)+\dots\nonumber\\
{\bf B}^\star(f_\sigma,\mathbf{E}_L) &= 	{\bf B}^\star_{0}(f_\sigma,\mathbf{E}_L) + \f{1}{c}{\bf B}^\star_{1}(f_\sigma,\mathbf{E}_L)+\dots\label{fastexpansion}
\end{align}

Using~\eqref{fastexpansion} we can cast equations~\eqref{Emaxwell}~and\eqref{BFaraday} as
\begin{equation}\label{Emymaxwell}
\nabla\times{\bf B}^\star = \f{4\pi}{c}\, {\bf J}_T +\f{1}{c} D_f {\bf E}^\star_{T}\lf[\dot f \ri] + \f{1}{c} D_{{\bf E}_L} {\bf E}^\star_{T}\lf[\dot {\bf E}_L \ri]
\end{equation}
and
\begin{equation}\label{BmyFaraday}
-\nabla\times{\bf E}^\star_T =\f{1}{c} D_f {\bf B}^\star\lf[\dot f \ri] + \f{1}{c} D_{{\bf E}_L} {\bf B}^\star\lf[\dot {\bf E}_L \ri].
\end{equation}
We find trivially that in the zeroth order expansion with respect to $1/c$
\begin{equation}
    \nabla \times {\bf E}_{T0}^{\star}=0,\quad \nabla \times {\bf B}_{0}^{\star}=0,
\end{equation}
which implies
\begin{equation}
{\bf E}^\star_{T0}= 0,\quad  {\bf B}^\star_{0} =0,
\end{equation}
because $\mathbf{E}_T$ and $\mathbf{B}$ are solenoidal.
Notice that the first three non-zero terms in the expansion of ${\bf J}$ are 
\begin{eqnarray}\label{jexpansion}
    {\bf J}_{0}&=&\sum_\sigma\frac{e_\sigma}{m_\sigma} \int d^{3} p\, f_\sigma {\bf p}, \quad {\bf J}_{2}=-\sum_\sigma\frac{e_\sigma}{2 m^{3}_\sigma} \int d^{3} p\, f_\sigma p^{2} {\bf p}, \label{Jexpansion}\\
    {\bf J}_{4}&=&\sum_\sigma\frac{3 \,e_\sigma}{8 m^{5}_\sigma} \int d^{3} p\, f_\sigma p^{4} {\bf p}\nonumber,\quad \dots
\end{eqnarray}
The absence of transverse fields to this order suggests absence of light waves. Below we will show that if we prepare plasma in this special state, light waves are going to be suppressed in future. In the first order
\begin{equation}
    \nabla \times {\bf E}_{T1}^{*}=0,\quad \frac{1}{c} {\nabla} \times {\bf B}_{1}^{*}=-\frac{4 \pi }{c} \nabla \times \Delta^{-1} \nabla\times  {\bf J}_{0}
\end{equation}
Thus the following (Darwin) magnetic field is recovered 
\begin{equation}\label{b1star}
{\bf B}^\star_{1} =- 4\pi \,\Delta^{-1} \nabla\times {\bf J}_{0}:= - \Delta^{-1}\nabla\times\sum_\sigma \f{4\pi e_\sigma}{m_\sigma}\int d^3 p \, f_\sigma {\bf p},
\end{equation}
In many circumstances we prefer to work with the transverse vector potential instead of the magnetic field:
\begin{equation}\label{A1star}
{\bf A}^\star_{T1} = -4\pi\, \Delta^{-1} \Pi_T\, {\bf J}_{0} , 
\end{equation}
where special care is taken to ensure that the Green's function acts on an object that is already in the transverse space. Meanwhile it turns out that the electric field is absent in this order
\begin{equation}\label{E1star}
{\bf E}^\star_{1} = 0
\end{equation}

This limit constitutes the well known Darwin approximation. As stated earlier it is typically applied in quasistatic situations. However \citet{Eremin_2013} show its utility in the intermediate range for high-frequency capacitively coupled discharges. The Darwin approximation offers a simple and efficient way of carrying out electromagnetic simulations as it removes the Courant condition plaguing explicit electromagnetic algorithms and can be implemented as a straightforward modification of electrostatic algorithms.

From~\eqref{b1star} and the $1/c^2$ expansion of~\eqref{Emaxwell} it is evident that this time
\begin{equation}\label{B2star}
{\bf B}^\star_{2} = 0
\end{equation}
On the other hand, from $1/c^2$ expansion of~\eqref{BmyFaraday} we see
\begin{equation}\label{estar2formal}
\nabla\times {\bf E}^\star_{T2} = - D_f {\bf B}^\star_{1} \lf[\dot f^{0} \ri] \equiv -\sum_\sigma D_{f_\sigma} {\bf B}^\star_{1} \lf[\dot f^{0}_\sigma \ri],
\end{equation}
where $D$ is a Fr\'{e}chet derivative and we only need to keep zeroth order terms (with respect to $1/c$ expansion) in the expression for $\dot f$ (see~\eqref{maxwellvlasov}). Because of~\eqref{gammafactor} it is clear that the effects of relativity are not felt in this limit and thus we have ${\bf v} \approx {\bf p}/m$. Thus, to this order,
\begin{equation}
    \dot{f}_\sigma^{0}+\frac{{\bf p}}{m_\sigma} \cdot {\nabla} f_\sigma+e_\sigma {\bf E}_{L} \cdot \frac{\partial f_\sigma}{\partial {\bf p}}=0
\end{equation}
From~\eqref{estar2formal} one obtains after  integration by parts
\begin{equation}\label{E2star}
{\bf E}^\star_{T2} = {-} \sum_\sigma\Delta^{-1} \Pi_T\lf(4\pi e_\sigma\, \nabla\cdot {\bf T}_{0\sigma} - \omega_{p\sigma}^2   {\bf E}_L  \ri),
\end{equation}
where we have introduced the zeroth order stress tensor
\begin{equation}
{\bf T}_{0\sigma} := \int d^3 p\, \f{{\bf p \, p}}{m^2_\sigma} f_\sigma,\label{Ttensor}
\end{equation} 
and plasma frequency
\begin{equation}
\omega_{p\sigma}^2  := \f{4\pi e^2_\sigma}{m_\sigma}\int d^3 p \,f_\sigma.\quad {\omega_p^2 := \sum_\sigma \omega_{p\sigma}}\label{plasmafreqoper}
\end{equation}
Note that due to the spatial dependence of $\omega_{p\sigma}^2$ the last term in~\eqref{E2star} does not vanish despite the fact that it is acted upon by the transverse projection operator.

 According to Eq.\,\eqref{E2star}, stress generates transverse electric fields.  Obviously, in quasineutral plasma if the species are moving with the same velocity and the distribution function, the term would vanish. Thus we conclude that it has a two-fluid nature. 
 {The origin of this term can be, in fact, traced back to the Ohm's law found in Braginskii multifluid model with quasineutrality. Similar reformulation leading to this term has been proposed for Maxwell-Euler system by~\citet{DEGOND2017467}. }
{A similar effect may be found in certain solids. In solids piezoelectric effect is the induction of an electric charge in response to an applied mechanical strain, see~\citet{ZHANG2014415}, for example. It has numerous applications, most notably in microphones.}

We can show that the stress-tensor term can be simplified for a non-uniform drifting Maxwellian (since only zeroth order corrections are required we can use the non-relativistic expression)
\begin{equation}
f({\bf r},{\bf v}) = n({\bf r}) \lf( \f{m}{\pi T} \ri)^{\f{3}{2}} e^{-\f{m({\bf v} - {\bf V({\bf r})})^2}{2 T}}.
\end{equation}
We have 
\begin{equation}
\int d^3 v\, v_j v_k f = \delta_{jk} \f{n T}{2 m} + n V_j V_k
\end{equation}
Next, because of the longitudinal form of the first term it vanishes under the transverse projection. The stress term survives, in general. Of course, the second term in~\eqref{E2star} vanishes in case of a uniform density.

The second term in~\eqref{Ttensor} demonstrates the coupling between the longitudinal mode, originating in Langmuir oscillations, and the transverse mode. This effect is akin to the generation of transverse waves from the longitudinal ones. However, in this approximation the field that is generated is of the static nature, rather than radiation.

To see what happens further we go to the next order. From the application of~\eqref{B2star} and $\dot{f}_{1} = 0$ to~\eqref{Emymaxwell} it is easy to see
\begin{equation}\label{E3star}
{\bf E}_{T3}^\star = 0.
\end{equation}
On the other hand, we have a rather more complicated correction to the magnetic field determined by
\begin{equation}
\nabla\times{\bf B}_{3}^\star = 4\pi {\bf J}_{T2}\lf[f \ri] + D_f {\bf E}^\star_{T2}\lf[\dot f^{0} \ri] + D_{{\bf E}_L} {\bf E}^\star_{T2}\lf[\dot {\bf E}_L^{0} \ri],
\end{equation}
which using~\eqref{Emaxwell2} and~\eqref{Jexpansion} leads to 
\begin{eqnarray}
{\nabla} \times {\bf B}_{3}^{*}&=&-\sum_\sigma\frac{4 \pi e_\sigma}{2 m^{3}_\sigma} \Pi_{T} \int d^{3}p\, f_\sigma p^{2} {\bf p}{-}\sum_\sigma\frac{4 \pi e_\sigma}{m_\sigma} \Delta^{-1}{\Pi}_{T}\nonumber \\ 
&\cdot&\int d^{3} {p} \, \Big(4\pi e_\sigma f_\sigma {\Pi}_{L} {\bf J}_{0}
+e_\sigma {\bf E}_{L}\left[\frac{{\bf p}}{m_\sigma} \cdot {\nabla} f_\sigma+{\bf E}_L \cdot\frac{\partial f_\sigma}{\partial {\bf p}}\right]-\nonumber\\
&-&{\nabla} \cdot\left[\frac{{\bf p}{{\bf p}}}{m_\sigma}\left(\frac{{\bf p}}{m_\sigma} \cdot { \nabla} f_\sigma +e_\sigma {\bf E}_{L} \cdot \frac{\partial f_\sigma}{\partial {\bf p}}\right)\right]\Big)
\end{eqnarray}
Thus we collect these terms and use integration by parts to obtain
\begin{eqnarray}\label{B3star}
{\bf A}_{T3}^\star &=&  {-}\Delta^{-2}\Pi_T\lf[ \nabla\cdot\nabla\cdot {\bf Q}_0 - \nabla\cdot \lf({\bf E}_L\;  {\bf W}  +   {\bf W} \;{\bf E}_L \ri)  -{\bf E}_L \nabla\cdot {\bf W} \ri]\nonumber\\
&-& 4\pi\Delta^{-1} \Pi_T{\bf J}_{2}  {+}\sum_\sigma  4\pi \Delta^{-2}\Pi_T\omega_{p\sigma}^2\Pi_L {\bf J}_{0} ,
\end{eqnarray}
where we have defined the charge-weighted heat flux tensor:
\begin{equation}
{\bf Q}_0 := \sum_\sigma  4\pi e_\sigma \int d^3p \, \f{{\bf p\, p\, p}}{m^3_\sigma} f_\sigma
\end{equation}
and introduced the velocity weighted by the plasma frequency
\begin{align}
\bm{W} := \sum_\sigma \frac{4\pi e_\sigma^2 }{m_\sigma}\int \frac{\bm{p}}{m_\sigma}f_\sigma\,d^3p.
\end{align}
It is somewhat remarkable that a {charge density weighted} heat flux produces a magnetic field in a dark plasma. {To our knowledge this has not been reported in the literature yet.}

\subsection{Stability of the dark slow manifold\label{dark_stability_sec}}
Our analysis of the slow manifold so far has been entirely based on asymptotic expansions. Therefore it is unclear how long an initially-dark Vlasov-Maxwell solution will remain dark. Any thorough analysis of this question will require detailed functional analysis. However, we will now show that arguments from \cite{kristiansen16} may be applied formally to the Vlasov-Maxwell system in order to obtain a heuristic that suggests the timescale is at least $O(c)$ as $c\rightarrow\infty$ for initial conditions prepared ``optimally close'' to the slow manifold. Thus, we believe there is compelling evidence to conjecture that darkness persists over large time intervals in non-relativistic plasmas.

\cite{kristiansen16} study slow manifolds that arise in a certain class of Hamiltonian fast-slow systems. The system phase space is assumed to be of the form $\mathcal{P} = \mathcal{X}\times\mathcal{Y}\ni (x,y)$, where $x$ is the slow variable, $y$ is the fast variable, and both $\mathcal{X}$ and $\mathcal{Y}$ are real Hilbert spaces. The fast space $\mathcal{Y}$ is allowed to be infinite dimensional, but the dimension of the slow space $\mathcal{X}$ is assumed to be finite. While the rigorous arguments in this work use finite-dimensionality of $\mathcal{X}$ in an essential way, at a formal level the arguments apply to infinite-dimensional $\mathcal{X}$ as well. The Poisson tensor on $\mathcal{P}$ is given by $\mathcal{J} = \epsilon\,\mathcal{J}_{\mathcal{X}} + \mathcal{J}_{\mathcal{Y}}$, where $\epsilon\ll 1$ and $\mathcal{J}_{\mathcal{X}},\mathcal{J}_{\mathcal{Y}}$ are Poisson tensors on $\mathcal{X}$ and $\mathcal{Y}$, respectively. The system Hamiltonian is required to be of the form 
\begin{align}
    H_\epsilon(x,y) = h_\epsilon(x) + \langle R_\epsilon(x),y \rangle+\frac{1}{2}\langle y,(L+a_\epsilon(x))y \rangle + \gamma_\epsilon(x,y),\label{kw_ham}
\end{align}
where all $\epsilon$-dependence is continuous, $L,a_\epsilon(x)$ are self adjoint, $L+a_\epsilon(x)$ is invertible, and $\gamma_\epsilon(x,y) = O(||y||^3)$. It is also assumed that $h_\epsilon(x),R_\epsilon(x),a_\epsilon(x)$ and $\gamma_\epsilon(x,y)$ are real analytic. In this context, $y=0$ defines a limiting slow manifold.

The first major result from \cite{kristiansen16}, Theorem 2.1, states that there is a near-identity canonical transformation $(x,y)\mapsto (\overline{x},\overline{y})$ that transforms the system Hamiltonian \eqref{kw_ham} into
\begin{align}
    \overline{H}_\epsilon(\overline{x},\overline{y}) = \overline{h}_\epsilon(\overline{x}) + \langle \overline{R}_\epsilon(\overline{x}),\overline{y} \rangle+\frac{1}{2}\langle \overline{y},(L+\overline{a}_\epsilon({\overline{x}}))\overline{y} \rangle + \overline{\gamma}_\epsilon(\overline{x},\overline{y}),
\end{align}
where $\overline{a}_\epsilon(\overline{x})$ is self-adjoint, all barred quantities are close to their un-barred counterparts, and, crucially, $\overline{R}_\epsilon = O(\exp(-\kappa/\epsilon))$ for some positive constant $\kappa$. The evolution equation for $\overline{y}$ is therefore $\dot{\overline{y}} = \mathcal{J}_{\mathcal{Y}}\,\nabla_{\overline{y}}\overline{H}_\epsilon = \mathcal{J}_{\mathcal{Y}}(L+\overline{a}_\epsilon(x))\overline{y} + \mathcal{J}_{\mathcal{Y}}\,\nabla_{\overline{y}}\overline{\gamma}_\epsilon + \mathcal{J}_{\mathcal{Y}}\overline{R}_\epsilon(\overline{x}) $. Since the right-hand-side of this equation is exponentially small at $\overline{y}=0$, the set $\{(\overline{x},\overline{y})\mid \overline{y}=0\}$ defines an exponentially-accurate slow manifold.

The barred coordinate system is constructed by applying a sequence of $N$ near-identity canonical transformations such that, after the $n^{\text{th}}$ ($n\leq N$) step in the iteration, the transformed $R_\epsilon$ is smaller by a factor of $\epsilon/\xi_n$ relative to the previous iteration. Here $\xi_n$ is a parameter that controls the loss of regularity introduced by the $n^{\text{th}}$ transformation; smaller $\xi$ corresponds to a smaller loss of regularity, but a less dramatic reduction in the size of the transformed $R_\epsilon$. In order to produce an exponential reduction in the size of the transformed $R_\epsilon$ while avoiding a catastrophic loss of regularity, the authors choose $\xi_n \sim 2\epsilon$ for large $n$ (so $\epsilon/\xi_n  \sim 1/2$) and $N\sim 1/\epsilon$. These choices ensure that size of the transformed $R_\epsilon$ is halved after a single iteration, and reduced by a factor of $2^{-1/\epsilon} = \exp(-[\text{ln}\,2]/\epsilon)$ after $N$ iterations, which is the desired exponential effect.

The next major result from \cite{kristiansen16}, Corollary 2.2, uses the special coordinates described in the previous paragraph to estimate the stability timescale of the $\overline{y}=0$ slow manifold. Assuming $L+a_\epsilon(x)$ is positive definite, the result states that a solution of the fast-slow system with $\overline{y}(\tau=0)=0$ will satisfy $\overline{y}(\tau) = O(\exp(-\kappa_1/\epsilon))$ for times $\tau\in[0,\kappa_2/\epsilon^2]$. In other words, if a solution starts on the exponentially-accurate slow manifold then it will remain exponentially close to that manifold on a timescale that is at least $O(\epsilon^{-2})$ in $\tau$. The proof is a Lyapunov-type argument that exploits the near-constancy of the function
\begin{align}
    I = \tfrac{1}{2}\langle \overline{y},(L+\overline{a}_\epsilon(\overline{x}))\overline{y}\rangle+\overline{\gamma}_\epsilon(\overline{x},\overline{y})
\end{align}
along solutions of the fast-slow system. Near invariance of $I$ follows from the following simple calculation:
\begin{align}
    \dot{I} &= \bigg\langle \mathcal{J}_{\mathcal{Y}}(L+\overline{a}_\epsilon(\overline{x}))\overline{y} + \mathcal{J}_{\mathcal{Y}}\,\nabla_{\overline{y}}\overline{\gamma}_\epsilon + \mathcal{J}_{\mathcal{Y}}\overline{R}_\epsilon(\overline{x}) ,(L+\overline{a}_\epsilon(\overline{x}))\overline{y}+\nabla_{\overline{y}}\overline{\gamma}_\epsilon\bigg\rangle+O(\epsilon)\nonumber\\
    & = \bigg\langle \mathcal{J}_{\mathcal{Y}}\overline{R}_\epsilon(\overline{x}) ,(L+\overline{a}_\epsilon(\overline{x}))\overline{y}+\nabla_{\overline{y}}\overline{\gamma}_\epsilon \bigg\rangle+O(\epsilon)\nonumber\\
    & = O(\exp(-\kappa/\epsilon)) + O(\epsilon).
\end{align}

The arguments from \cite{kristiansen16} apply formally to the Vlasov-Maxwell system because of the following. First, we recall that the Vlasov-Maxwell system may be written as a Hamiltonian system on the space $\mathcal{P}=\mathcal{X}\times\mathcal{Y}$, where $\mathcal{X}\ni f$ is the space of canonical momentum distribution functions $f(\mathbf{x},\bm{\pi})$, and $\mathcal{Y}\ni (\mathbf{E}_T,\mathbf{A}_T)$ comprises pairs of transverse vector potentials $\mathbf{A}_T$ and transverse electric fields $\mathbf{E}_T$. The Poisson bracket of functionals $F(f,\mathbf{E}_T,\mathbf{A}_T),G(f,\mathbf{E}_T,\mathbf{A}_T)$ is given by $\{F,G\} = \epsilon\,\{F,G\}_{\mathcal{X}}+\{F,G\}_{\mathcal{Y}}$, where $\epsilon = 1/c$ and 
\begin{align}
    \{F,G\}_{\mathcal{X}} & = \int \left[\frac{\delta F}{\delta f},\frac{\delta G}{\delta f}\right]f\,d^3\mathbf{r}\,\,d^3\bm{\pi}\nonumber\\
    \{F,G\}_{\mathcal{Y}} & = 4\pi\int \left(\frac{\delta F}{\delta\mathbf{E}_T}\cdot\frac{\delta G}{\delta\mathbf{A}_T} -\frac{\delta G}{\delta\mathbf{E}_T}\cdot\frac{\delta F}{\delta\mathbf{A}_T} \right) \,d^3\mathbf{r}.
\end{align}
Here $[\cdot,\cdot]$ denotes the usual canonical Poisson bracket on $(\mathbf{x},\bm{\pi})$-space, as appropriate when working with canonical momenta $\bm{\pi}$. { Note that this is the only section where we work in coordinates which depend on canonical momentum for consistency with~\cite{kristiansen16}.} This bracket is obtained by quotioning the ``canonical'' Vlasov-Maxwell bracket discussed in \cite{Marsden_Weinstein_1982} by gauge transformations.
The system Hamiltonian is
\begin{align}
    H_\epsilon(x,y) &= \frac{1}{8\pi}\int \mathbf{E}_T\cdot\mathbf{E}_T\,d^3\mathbf{r} + \frac{1}{8\pi}\int \nabla\times\mathbf{A}_T\cdot\nabla\times\mathbf{A}_T\,d^3\mathbf{r} \nonumber\\
    &+ \frac{1}{8\pi}\int \mathbf{E}_L(f)\cdot\mathbf{E}_L(f)\,d^3\mathbf{r} + \int \frac{1}{2m} \left(\bm{\pi}-\epsilon\,e\mathbf{A}_T\right)^2\,f\,d^3\mathbf{r}\,d^3\bm{\pi},
\end{align}
where the longitudinal electric field $\mathbf{E}_L(f) = -\nabla\varphi(f)$ is the unique solution of the elliptic partial differential equation $-\Delta \varphi(f) =4\pi e\,n(f) - 4\pi \rho_0$, where $\rho_0$ denotes a constant neutralizing background charge and $n(f) = \int f(\mathbf{r},\bm{\pi})\,d^3\bm{\pi}$ is the number density. Note that the time variable $\tau$ for this Hamiltonian formulation of Vlasov-Maxwell is related to the time variable $t$ used elsewhere in this Article by $\tau = c \,t$. As required by \cite{kristiansen16}, the Poisson bracket has a product structure that is compatible with the fast-slow split, and the Hamiltonian has the form \eqref{kw_ham}, with
\begin{align}
    L(\mathbf{E}_T,\mathbf{A}_T) & = \tfrac{1}{4\pi}(\mathbf{E}_T,-\Delta\,\mathbf{A}_T)\\
    a_\epsilon(f)(\mathbf{E}_T,\mathbf{A}_T) & = \epsilon^2\,(0,\tfrac{e^2\,n(f)}{m}\,\mathbf{A}_T)\\
    h_\epsilon(f) & = \int \frac{1}{2m} |\bm{\pi}|^2\,f\,d^3\mathbf{r}\,d^3\bm{\pi}+ \frac{1}{8\pi}\int \mathbf{E}_L(f)\cdot\mathbf{E}_L(f)\,d^3\mathbf{r} \\
    R_\epsilon(f) & = -\epsilon\,e\int \tfrac{\bm{\pi}}{m}\,f\,d^3\bm{\pi}\\
    \gamma_\epsilon(f,\mathbf{E}_T,\mathbf{A}_T) & = 0.
\end{align}
It follows that the formal transformation {$(x,y)\mapsto (\overline{x},\overline{y})$, mentioned earlier,} that ``flattens" the slow manifold can be derived for this system much as in \cite{kristiansen16}. Analyticity of the Vlasov-Maxwell system in $(f,\mathbf{E}_T,\mathbf{A}_T)$ then suggests that optimal truncation of the formal transformation should produce an exponentially-accurate slow manifold, as in Theorem 2.1 from \cite{kristiansen16}. If this is indeed the case, then positive-definiteness of $L+a_\epsilon$ (easily seen to be true) together with the proof of Corollary 2.2 in \cite{kristiansen16} implies normal stability of the exponentially-accurate dark slow manifold for $\tau\in[0,c_2/\epsilon^2]$, or {equivalently} $t\in[0,c_2/\epsilon]$. In other words, there exist solutions of the Vlasov-Maxwell system with initially exponentially-small light-wave activity that remain exponentially close to the dark slow manifold on $O(1/\epsilon)=O(c)$ time intervals.

While this argument suggests a concrete path toward proving persistence of dark plasma states in the Vlasov-Maxwell phase space on $O(1/\epsilon)$ timescales, it is important to understand the argument's shortcomings. First and foremost, in \cite{kristiansen16}, it was crucial for the authors to understand precisely the loss of regularity introduced by each step in their sequence of canonical transformations. Due to specifics of the function-analytic setting considered by \cite{kristiansen16}, the details of this regularity loss may very well differ for the Vlasov-Maxwell system. Any modifications that might arise need to be studied and accounted for appropriately in the optimal truncation procedure. 
Finally, this argument establishes long-term stability of the optimally-truncated slow manifold, but not of lower-order truncations. Stability of these lower-order truncations must be assessed using other methods.

On the other hand, the issues inherent to this argument disappear if the Vlasov-Maxwell system is replaced with a continuous-time structure-preserving discretization as in \cite{kraus16} or \cite{burby_2017}. It would be interesting to use this argument to study non-relativistic slow manifolds in discrete Vlasov-Maxwell systems in the future.


\section{{Dark slow manifold} dynamics as a Hamiltonian system\label{Dark plasma dynamics as a Hamiltonian system}}
So far we have established the existence of a formal slow manifold in the infinite-dimensional Vlasov-Maxwell phase space on which light waves are inactive. If a plasma's initial state is prepared to lie on this slow manifold, the ensuing plasma motion will not emit light for some time, and in this sense will be dark. We deduced a heuristic lower bound on the timescale over which darkness persists for optimally-dark initial conditions in Section \ref{dark_stability_sec}. In this section we will deduce the dynamical equations that govern dynamics on the dark slow manifold, as well as their Hamiltonian structure.

In general, given a fast-slow system $\epsilon\,\dot{y}=f_\epsilon(x,y)$, $\dot{x} = g_\epsilon(x,y)$ with formal slow manifold $y_\epsilon^*$, dynamics on the slow manifold is governed by the system of equations $\dot{x} = g_\epsilon(x,y_\epsilon^*(x))$, which may be interpreted as a closure of the evolution equations for the slow variables. While the right-hand-side of this evolution equation $g_\epsilon(x,y_\epsilon^*(x))$ is unwieldy since it is a formal power series, it can be truncated at any finite order in a straightforward, if tedious manner. Such truncations provide approximate descriptions of the slow dynamics.

The above method of describing slow manifold dynamics can be applied to dark plasmas. However, doing so would ignore the fact that the dark slow manifold sits inside of the Vlasov-Maxwell system, which is known to have a Hamiltonian structure (see \citet{morrison1980} and \citet{Marsden_Weinstein_1982}). As discussed in \citet{mackay04} and \citet{burby2020}, when a slow manifold arises in a Hamiltonian system, the slow dynamics naturally inherits a Hamiltonian structure of its own. Therefore a better method for describing 
{such} dynamics is to derive this induced Hamiltonian structure, which in general comprises a Hamiltonian functional and a Poisson bracket. Once this structure has been identified, dynamics on the slow manifold may be recovered from Hamilton's equations in Poisson bracket form. The benefit of this approach is that it enables truncating the slow dynamics while preserving the underlying Hamiltonian structure. In contrast, the naive truncation procedure above breaks the Hamiltonian structure in general.

Since the dark slow manifold depends on $\epsilon$, the Hamiltonian and Poisson bracket induced on the slow manifold must also depend on $\epsilon$. Therefore expanding and truncating the Poisson bracket form of the slow evolution equations in powers of $\epsilon$ necessarily involves power series expansion and truncation of both the bracket and the Hamiltonian. This complicates the task of developing Hamiltonian approximations of the slow dynamics since a truncated power series expansion of the Poisson bracket need not satisfy the Jacobi identity. (This follows from the quadratic dependence of the Jacobi identity on the bracket.) To overcome this difficulty, we will apply the method introduced in \cite{burby2017magnetohydrodynamic} for describing the bracket on the slow manifold underlying the magnetohydrdynamic equations. In particular, we will apply a near-identity non-canonical transformation to the slow manifold that causes the transformed Poisson bracket to have a simple closed-form expression. We say that the transformation \emph{rectifies} the bracket. We will then define our approximate slow evolution equations in the transformed variables by using the full transformed Poisson bracket, but a truncated transformed Hamiltonian. In this way, we ensure that our approximate slow evolution equations comprise a genuine Hamiltonian system.

In the remainder of this section, we will (\ref{Mathematical preliminaries}) supply some necessary mathematical background, (\ref{Why is dark plasma dynamics Hamiltonian?}) explain how and why the slow manifold inherits a Hamiltonian structure, (\ref{Derivation of the rectifying transformation}) derive the near-identity transformation that rectifies the slow-manifold Poisson bracket, and finally (\ref{Calculation of the dark Poisson bracket and post-Darwin Hamiltonian}) derive the transformed slow manifold Hamiltonian to the first post-Darwin order. In this manner, we will provide the first dynamical description of dark plasma dynamics that extends beyond the Darwin approximation. We warn the reader that the discussion in this Section requires substantially more mathematical background than previous Section. To alleviate some of this additional complexity, we recommend consulting the review article by MacKay \cite{mackay_2020}, the excellent textbook by Abraham and Marsden \cite{Abraham_2008}, the lectures on symplectic geometry by da Silva \cite{Silva_book_2008}, as well as appendices \ref{inft_forms} and \ref{rect_exist_sec}.


\subsection{Mathematical Preliminaries\label{Mathematical preliminaries}}

First we will review some well-known results from finite dimensional Lagrangian mechanics. For details see~\citet{M98,CB09,jose98,marsden99}, for instance, although we follow slightly different conventions. The Lagrangian for an individual charged particle moving in an electromagnetic field may be written in the ``phase space" form as
\begin{equation}
L = \lf(\mathbf{p}+\frac{e}{c} \mathbf{A}\ri)\cdot {\bf \dot r} - \lf(e \,\Phi({\bf r}) + \f{m v^2}{2} \ri) =: {\theta}_B \cdot {\bf \dot z} - H({\bf z}),
\end{equation}
where we have introduced the Lagrange $1$-form $\theta_B =  \lf(\mathbf{p}+\f{e}{c}{\bf A}\ri)\cdot d\mathbf{r}$ and the phase space coordinate $\mathbf{z} = (\mathbf{r},\mathbf{p})$. {For review of differential forms for plasma physicist see~\citet{mackay_2020}. }
The symplectic $2$-form on the momentum phase space associated with $L$ can be obtained via ${\omega}_B = -d{\theta}_B$ {and has the form}

\begin{equation}
    {\omega_B = d{\bf r}\wedge d{\bf p} + \frac{e}{c} dr^i \wedge dr^j \,\pp{A_i}{r^j}}
\end{equation}
which leads to the symplectic version of Hamilton's equations {$\dot{q} = \partial H/\partial p$ and $\dot{p}= - \partial H/\partial q$}
\begin{equation}\label{finitehamilton}
{\omega_B({\bf \dot z}, \cdot)} \equiv i_{\bf \dot z} {\omega}_B = d H.
\end{equation}
The symplectic form can also be inverted to obtain the Poisson tensor 
\begin{equation}
\calJ_B = \begin{pmatrix}
0 && \delta^{ij} \\
-\delta^{ij}  && \epsilon_{ijk} \dfrac{e}{c} B^k
\end{pmatrix},\label{nearcocanonicalform}
\end{equation}
and thus the Poisson bracket is
\begin{equation}
\{f,g\} {:=\pp{f}{ z^\alpha} J^{\alpha\beta} \pp{g}{ z^\beta}}= \pp{f}{\bf r} \cdot\pp{g}{\bf p}  - \pp{g}{\bf r} \cdot\pp{f}{\bf p} + \f{e{\bf B}}{c}\cdot \pp{f}{{\bf p}}\times \pp{g}{{\bf p}}.
\end{equation}
{For example in coordinates one recovers $\dot z^\alpha = \{z, H\} = J^{\alpha\beta} \partial H/\partial z^\beta$.}
{When $\mathbf{B}=0$, we write $\theta_B = \theta_0 {={\bf p}\cdot {\bf dr}}$} and we recover the canonical symplectic form $\omega_0{=-d\theta_0 = d{\bf r}\wedge d{\bf p}}$ and Poisson tensor $\calJ_0$.

Next we generalize to treat a distribution of particles. Neglecting the electromagnetic field for now, the analogue of $\theta_B$ for an ensemble of charged particles, which we will denote $\Theta$, is a $1$-form on the infinite-dimensional space of Lagrangian configuration maps
\begin{equation}
g: \mathring{\calP} \rightarrow \calP,\label{eulerlagrangemap}
\end{equation}
where $\mathring{\calP}\ni (\mathbf{r}_0,\mathbf{p}_0)$ denotes the space of particle labels and $\calP\ni (\mathbf{r},\mathbf{p})$ denotes the Eulerian phase space. {(We remark that, as sets, $\mathring{\calP}$ and $\calP$ are the same. The notational distinction reflects the different physical interpretations of particle locations and particle labels.)} In particular, using the notation introduced in Appendix \ref{inft_forms}, we have
\begin{align}
\iota_{\dot{g}}\Theta_g = \int \iota_V\theta_0\,f\,d^3r\,d^3p,
\end{align}
where the Eulerian phase space velocity is given by $V = \dot{g}\circ g^{-1}$ and $f$ is defined in terms of $g$ according to
\begin{align}
f\,d^3r\,d^3p = g_{*}(\mathring{f}\,d^3r_0\,d^3p_0).\label{pushforwardf0_new}
\end{align} 
Here $\mathring{f}$ is some fixed reference distribution on label space. 

To find the symplectic form associated with $\Theta$, we must compute the exterior derivative $\Omega = -d\Theta$. We proceed by applying the formula \eqref{d_one_form} from Appendix \ref{inft_forms} with $\alpha = \Theta$. To begin, we introduce the functional $I$ on the space of curves $g(t)$ with $g(-1)$ and $g(1)$ fixed whose value at $g(t)$ is $I(g) = \int_{-1}^{1}\int \iota_V\theta_0\,f\,d^3r\,d^3p\,dt$. The first variation of $I$ is given by
\begin{align}
    \delta I(g)[\delta g] = \int_{-1}^{1}\int \iota_V\iota_\xi d\theta_0\,f\,d^3r\,d^3p\,dt,
\end{align}
where $\xi = \delta g\circ g^{-1}$ and we have used the identities $\delta V = \dot{\xi} + \mathcal{L}_{V}\xi$ and $\delta (f\,d^3r\,d^3p) = -\mathcal{L}_{\xi}(f\,d^3r\,d^3p)$, as in the Euler-Poincar\'e theory developed by \citet{HOLM19981}. Now introducing $g(t) $ and $\delta g(t)$ as in Appendix \ref{inft_forms}, the formula \eqref{d_one_form} implies
\begin{align}
    \iota_{\delta g_2}\iota_{\delta g_1}\Omega_g & = -\lim_{a\rightarrow 0}\frac{1}{2a} \delta I(g)[\mathbb{I}_{[-a,a]}\delta g]\nonumber\\
    & =  -\lim_{a\rightarrow 0}\frac{1}{2a}\int_{-a}^{a}\int \iota_{V}\iota_{\xi} d\theta_0\,f\,d^3r\,d^3p\,dt\nonumber\\
    & = -\int \iota_{\xi_2}\iota_{\xi_1} d\theta_0\,f\,d^3r\,d^3p.
\end{align}

To invert the $2$-form $\Omega$ and obtain the Poisson bracket $\{\cdot,\cdot\}$ on the space of Lagrangian configuration maps, we first compute the general Hamiltonian vector field $X_F = \dot{g}_F = V_F\circ g$ associated with a real-valued Hamiltonian functional $F(g)$. By definition, $X_F$ satisfies 
\begin{align}
\iota_{X_F}\Omega= dF.\label{infinitehamequations}
\end{align} 
See Appendix \ref{inft_forms} for a definition of exterior derivative of a scalar functional. In particular, if $W = \delta g = \xi\circ g$ is any vector on $g$-space then we must have $\iota_W\iota_{X_F}\Omega = \iota_W dF$. After defining the functional derivative $\delta F/\delta g$ according to
\begin{align}
\iota_WdF = \int \frac{\delta F}{\delta g}\cdot \delta g\,\mathring{f}\,d^3r_0\,d^3p_0,
\end{align}
 such that $\delta F/\delta g$ is a $1$-form at $g(\mathbf{r}_0,\mathbf{p}_0)$, we therefore infer that Hamilton's equation for $X_F$ is equivalent to 
\begin{align}
\int \iota_{\xi}\iota_{V_F}\omega_0\,f\,d^3r\,d^3p = \int \iota_{\xi}\left(\frac{\delta F}{\delta g}\circ g^{-1}\right)\,f\,d^3r\,d^3p,
\end{align}
for each vector field $\xi$ on the Eulerian phase space. Since $f$ is positive everywhere, we conclude $X_F$ is determined by
\begin{align}\label{VlasovPoissonVF}
V_F = \mathcal{J}_0\cdot \left(\frac{\delta F}{\delta g}\circ g^{-1}\right),
\end{align}
where $\mathcal{J}_0$ is the Poisson tensor associated with $\omega_0$. The following manipulation now immediately gives the Poisson bracket:
\begin{align}
\{G,F\} &= \mathcal{L}_{X_F}G\nonumber\\
&= \int \frac{\delta G}{\delta g}\cdot \dot{g}_F\,\mathring{f}\,d^3r_0\,d^3p_0\nonumber\\
&= \int \left(\frac{\delta G}{\delta g}\circ g^{-1}\right)\cdot\mathcal{J}_0\cdot\left(\frac{\delta F}{\delta g}\circ g^{-1}\right)\,f\,d^3r\,d^3p.\label{Lagrangian_bracket}
\end{align}
Notice that if $F(g) = \mathcal{F}(g_*\mathring{f}\,d^3r_0\,d^3p_0)$ and $G(g) = \mathcal{G}(g_*\mathring{f}\,d^3r_0\,d^3p_0)$ for some real-valued functionals $\mathcal{F},\mathcal{G}$ on the space of volume forms then their Poisson bracket is given by
\begin{align}
\{G,F\}(g) = \int d\frac{\delta \mathcal{G}}{\delta f}\cdot\mathcal{J}_0\cdot d\frac{\delta \mathcal{F}}{\delta f}\,f\,d^3r\,d^3p = \{\mathcal{F},\mathcal{G}\}_{LP}(g_* \mathring{f}\,d^3r_0\,d^3p_0),\label{poisson_property}
\end{align}
where $\{\cdot,\cdot\}_{LP}$ is the well-known Lie-Poisson bracket on the space of volume forms. Eq.\,\eqref{poisson_property} says that the mapping $g\mapsto g_*(\mathring{f}\,d^3r_0\,d^3p_0)$ is a Poisson map between the space of Lagrangian configuration maps with the bracket $\{\cdot,\cdot\}$ and the space of volume forms equipped with the Lie-Poisson bracket. We remark, however, that the image of this Poisson map is not equal to the space of volume forms, since pushforward preserves the total number of particles of $\mathring{f}$. 

We can repeat the above procedure for the full multi-species Vlasov-Maxwell system. The infinite-dimensional phase space now comprises tuples of the form $z=(g_\sigma, \mathbf{A},\mathbf{E})$, where $g_\sigma$ is the Lagrangian configuration map for species $\sigma$, $\bm{A}$ is the vector potential, and $\bm{E}$ is the electric field. The space of all such $z$ is denoted $Z$.
In this case the Lagrange one-form paired with the tangent vector $\dot z := (\dot{g}_\sigma, \dot{\mathbf{A}}, \dot{\mathbf{E}} )$ is
\begin{equation}\label{vlasovmaxwelllagrangeoneform}
    \iota_{\dot{z}}{\theta}_{\text{MV}\,z} = \sum_\sigma\int \iota_{V_\sigma}\theta_{B\sigma} { f}_\sigma\,d^3r\,d^3p - \f{1}{4\pi c}\int  {\bf E}\cdot \dot {\bf A} \,d^3r.
\end{equation}
The formula \eqref{d_one_form} with $\alpha = \theta_{\text{MV}}$ is then

implies that the symplectic form $\Omega_{\text{MV}} = -d\theta_{\text{MV}}$ is given by
\begin{align}
\iota_{\delta z_2}\iota_{\delta z_1}\Omega_{\text{MV}\,z}& =  \sum_\sigma\int \iota_{\xi_{\sigma2}}\iota_{\xi_{\sigma1}}\omega_{B\sigma}\,f_\sigma\,d^3r\,d^3p\nonumber\\
& +\frac{1}{4\pi c}\int \left(  \delta\mathbf{E}_1\cdot\delta{\mathbf{A}}_2-\delta{\mathbf{E}}_2\cdot\delta\mathbf{A}_1\right)\,d^3r\nonumber\\
& -\sum_\sigma \int \frac{e_\sigma}{c}\left(\delta\mathbf{A}_1\cdot \xi_{\sigma2}^{\mathbf{r}} - \delta{\mathbf{A}}_2\cdot \xi_{\sigma1}^{\mathbf{r}}\right)\,f_\sigma\,d^3r\,d^3p.\label{MV_symplectic_form}
\end{align}

To find the general Hamiltonian vector field $X_F$ associated with a Hamiltonian $F(g_\sigma,\mathbf{E},\mathbf{A})$, we first record the definition of the functional derivatives $\delta F/\delta g_\sigma$, $\delta F/\delta \mathbf{E}$, and $\delta F/\delta\mathbf{A}$:
\begin{align}
\iota_{\delta z}dF = \sum_\sigma \int \frac{\delta F}{\delta g_\sigma}\cdot \delta g_\sigma\,\mathring{f}_\sigma\,d^3r_0\,d^3p_0 + \int \frac{\delta F}{\delta \mathbf{E}}\cdot \delta\mathbf{E}\,d^3r + \int \frac{\delta F}{\delta \mathbf{A}}\cdot \delta\mathbf{A}\,d^3r.
\end{align}
Then we impose the condition $\iota_{\delta z}\iota_{X_F}\Omega_{\text{MV}} = \iota_{\delta z}dF$ for all vectors $\delta z$, which leads to the following expression for $X_F=(V_{F\sigma}\circ g_\sigma,\dot{\mathbf{E}}_F,\dot{\mathbf{A}}_F)$:
\begin{align}
V_{F\sigma} &= \mathcal{J}_{B\sigma}\cdot\left(\frac{\delta F}{\delta g_\sigma}\circ g_\sigma^{-1} - 4\pi e_\sigma \frac{\delta F}{\delta\mathbf{E}}\cdot d\mathbf{r}\right)\label{VdotH}\\
\dot{\mathbf{E}}_F& = -4\pi\sum_\sigma e_\sigma \int \left[\mathcal{J}_{B\sigma}\cdot\left(\frac{\delta F}{\delta g_\sigma}\circ g_\sigma^{-1}\right)\right]^{\mathbf{r}}\,f_\sigma\,d^3p + 4\pi c \frac{\delta F}{\delta\mathbf{A}}\label{EdotH}\\
\dot{\mathbf{A}}_F& = -4\pi c\frac{\delta F}{\delta \mathbf{E}}.\label{AdotH}
\end{align}
Finally, we again use the identity $\{G,F\}_{\text{MV}} = \mathcal{L}_{X_F}G = \iota_{X_F}dG$ to find that the Poisson bracket for the Vlasov-Maxwell system is given by
\begin{align}
&\{G,F\}_{\text{MV}} \nonumber\\
& = \sum_\sigma\int \left(\frac{\delta G}{\delta g_\sigma}\circ g_{\sigma}^{-1} - 4\pi e_\sigma \frac{\delta G}{\delta\mathbf{E}}\cdot d\mathbf{r}\right)\cdot\mathcal{J}_{B\sigma}\cdot\left(\frac{\delta F}{\delta g_\sigma}\circ g_{\sigma}^{-1} - 4\pi e_\sigma \frac{\delta F}{\delta\mathbf{E}}\cdot d\mathbf{r}\right)\,f_\sigma d^3r\,d^3p\nonumber\\
& + 4\pi c\int \left(\frac{\delta G}{\delta\mathbf{E}}\cdot\frac{\delta F}{\delta\mathbf{A}} - \frac{\delta G}{\delta\mathbf{A}}\cdot\frac{\delta F}{\delta \mathbf{E}}\right)\,d^3r\label{VM_bracket_canonical}
\end{align}
This bracket is a special case of Eq.\,(3.252) in \citet{Burby_thesis}. It is related to the well-known Morrison-Marsden-Weinstein bracket (see \citet{morrison1980} and \citet{Marsden_Weinstein_1982}) on the space of tuples $(f_\sigma,\mathbf{E},\mathbf{B})$ by the Poisson mapping
\begin{align}
(g_\sigma,\mathbf{E},\mathbf{A})\mapsto (g_{\sigma*}(\mathring{f}_\sigma\,d^3r_0\,d^3p_0),\mathbf{E},\nabla\times\mathbf{A}).
\end{align}
As is well-known, the Hamiltonian for the Vlasov-Maxwell system is the sum of particle and field energies,
\begin{align}
H_{\text{MV}}(g_\sigma,\mathbf{E},\mathbf{A}) = \sum_\sigma \int m_\sigma c^2\,\gamma_\sigma\,f_\sigma\,d^3r\,d^3p + \frac{1}{8\pi}\int (|\mathbf{E}|^2 + |\nabla\times\mathbf{A}|^2)\,d^3r,\label{MVHam}
\end{align}
where $\gamma_\sigma = \sqrt{1 + |\mathbf{p}|^2/(m_\sigma c)^2}$ is the Lorentz factor for species $\sigma$.


\subsection{Why is Dark dynamics Hamiltonian?\label{Why is dark plasma dynamics Hamiltonian?}}

We will now give a precise explanation for why plasma dynamics on the dark slow manifold possess a Hamiltonian structure. Further discussion of Hamiltonian structure on slow manifolds more generally is found in the review articles \citet{mackay04} and \cite{burby2020}. 

Suppose that $(Z,\Omega)$ is a symplectic manifold with symplectic form $\Omega$. If $\Lambda\subset Z$ is \emph{any} submanifold of $Z$ with inclusion map $I_\Lambda:\Lambda\rightarrow Z$ then $\Lambda$ inherits a  $2$-form $\Omega_\Lambda = I_\Lambda^*\Omega $, where $I_\Lambda^*$ denotes the pullback to $\Lambda$. This $2$-form is closed since $d\Omega_\Lambda = dI_\Lambda^*\Omega = I_\Lambda^*d\Omega = 0$, which means that $\Lambda$ is intrinsically a {presymplectic manifold}.\footnote{A presymplectic manifold is a smooth manifold equipped with a closed $2$-form. A symplectic manifold is a presymplectic manifold whose closed $2$-form is non-degenerate.} If $\Omega_\Lambda$ happens to be non-degenerate everywhere on $\Lambda$, then $\Lambda$ is intrinsically a symplectic manifold. As explained in \citet{Sniatycki_1974}, the Poisson bracket induced by $\Omega_\Lambda$ in the non-degenerate case coincides with the well-known Dirac bracket for constrained mechanical systems; the submanifold $\Lambda$ represents the constraint.

If $X_H$ is a Hamiltonian vector field on $(Z,\Omega)$ and $\Lambda$ is an invariant submanifold, then $X_H$ is tangent to $\Lambda$ and the induced vector field $X_{H\Lambda}$ on $\Lambda$ is Hamiltonian with respect to the presymplectic structure $\Omega_\Lambda$ discussed above. To see this, first note that since $X_H$ is Hamiltonian we have $\iota_{X_H}\Omega = dH$ in $Z$. Pulling back this equation to $\Lambda$ along $I_\lambda$ gives $\iota_{X_{H\Lambda}}\Omega_\Lambda = dH_\Lambda$ where $H_\Lambda = I_\Lambda^*H$, which says that $X_{H\Lambda}$ is a Hamiltonian vector field on $\Lambda$ with Hamiltonian $H_\Lambda$. In this sense, dynamics on invariant manifolds contained in symplectic manifolds always inherit their own intrinsic Hamiltonian structure.

As we explained in Section \ref{The Vlasov-Maxwell equations as a fast-slow system}, the Eulerian, gauge-invariant form of the Vlasov-Maxwell system contains a (formal) invariant manifold $S$ equal to the dark slow manifold. We constructed $S$ as the constraint set $\mathbf{E}_T = \mathbf{E}_T^\star(f_\sigma,\mathbf{E}_L)$, $\mathbf{B} = \mathbf{B}_T^\star(f_\sigma,\mathbf{E}_L)$ inside of $(f_\sigma,\mathbf{E},\mathbf{B})$-space. Since the Eulerian, gauge-independent phase space $(f_\sigma,\mathbf{E},\mathbf{B})$ is related to the Lagrangian, gauge-dependent phase space $(g_\sigma,\mathbf{E},\mathbf{A})$ by elimination of the gauge and relabeling degrees of freedom, we may ``undo" the gauge and relabeling symmetries to construct an invariant manifold $\widetilde{S}$ in $z=(g_\sigma,\mathbf{E},\mathbf{A})$-space that projects onto the dark slow manifold $S$. To wit, $\widetilde{S}\subset Z$ is defined by the constraints $\mathbf{A}_T = \mathbf{A}_T^{\star\star}(g_\sigma,\mathbf{E}_L)$ and $\mathbf{E}_T =\mathbf{E}_T^{\star\star}(g_\sigma,\mathbf{E}_L)$, where the slaving functions $\mathbf{A}_T^{\star\star}$ and $\mathbf{E}_T^{\star\star}$ are given by
\begin{align}
\mathbf{A}_T^{\star\star}(g_\sigma,\mathbf{E}_L) & = \mathbf{A}_T^{\star}(f_\sigma,\mathbf{E}_L)\\
\mathbf{E}_T^{\star\star}(g_\sigma,\mathbf{E}_L) & = \mathbf{E}_T^{\star}(f_\sigma,\mathbf{E}_L)\\
g_{\sigma *}(\mathring{f}\,d^3r_0\,d^3p_0) &=f_\sigma \,d^3r\,d^3p,\nonumber
\end{align}
and $\mathbf{A}_T^\star(f_\sigma,\mathbf{E}_L)$ is the unique transverse vector field on $Q$ whose curl is $\mathbf{B}^*(f_\sigma,\mathbf{E}_L)$.

Since we have already shown that the gauge-dependent, Lagrangian form of the Vlasov-Maxwell system is a Hamiltonian system on the symplectic manifold $(Z,\Omega_{\text{MV}})$, the preceding ramarks with $\Lambda = \widetilde{S}$ imply that dark plasma dynamics (in gauge-dependent, Lagrangian form) must possess an intrinsic Hamiltonian structure. As we will see, the $2$-form on $\widetilde{S}$ happens to be non-degenerate, which implies that the (all-orders) Poisson bracket on $\widetilde{S}$ is a Dirac bracket.

To compute this structure, we must find both the Hamiltonian $H_{\widetilde{S}}$ and the $2$-form $\Omega_{\widetilde{S}}$ on $\widetilde{S}$. To that end, we introduce the inclusion map $I_{\widetilde{S}}:(g_\sigma,\mathbf{E}_L,\mathbf{A}_L)\mapsto (g_\sigma,\mathbf{E},\mathbf{A})$ defined by
\begin{align}
\mathbf{E} &= \mathbf{E}_L  + \mathbf{E}_T^\star(f_\sigma,\mathbf{E}_L)\\
\mathbf{A} & = \mathbf{A}_L + \mathbf{A}_T^*(f_\sigma,\mathbf{E}_L).
\end{align}
Then we pull back the primitive $1$-form $\theta_{\text{MV}}$ for $\Omega_{\text{MV}}$ along $I_{\widetilde{S}}$ to obtain the Lagrange $1$-form $\theta^\star$ for dark plasma dynamics:
\begin{align}
\iota_{\dot{x}}\theta^\star_x& = \sum_\sigma \int \left(\mathbf{p}+\frac{e_\sigma}{c}\mathbf{A}_L + \frac{e_\sigma}{c}\mathbf{A}_T^\star\right)\cdot V_\sigma^{\mathbf{r}}\,f_\sigma\,d^3r\,d^3p\nonumber\\
&- \frac{1}{4\pi c}\int (\mathbf{E}_L\cdot \dot{\mathbf{A}}_L + \mathbf{E}_{T}^\star\cdot D_{\mathbf{E}_L}\mathbf{A}_T^\star[\dot{\mathbf{E}}_L])\,d^3r.\nonumber\\
&+ \frac{1}{4\pi c}\sum_\sigma\int ( \mathbf{E}_{T}^\star\cdot D_f\mathbf{A}_T^\star[\text{div}(V_\sigma f_\sigma)])\,d^3r\label{dadrk_Lagrange_form}
\end{align}
Here $x=(g_\sigma,\mathbf{E}_L)$ parameterizes the dark slow manifold and $V_\sigma,f_\sigma$ are defined in terms of $g_\sigma$ as in Section \ref{Mathematical preliminaries}. We also pull back the Vlasov-Maxwell Hamiltonian along $I_{\widetilde{S}}$ to obtain the Hamilton function $H^\star$ governing dark plasma dynamics:
\begin{align}
H^\star(x) & = \sum_\sigma \int m_\sigma c^2\,\gamma_\sigma\,f_\sigma\,d^3r\,d^3p + \frac{1}{8\pi}\int (|\mathbf{E}_L|^2 + |\mathbf{E}_T^\star|^2 + |\nabla\times\mathbf{A}_T^\star|^2)\,d^3r,
\end{align}
The all-orders evolution equations for a dark plasma are then $\dot{x} = X_{H^\star}(x)$, where 
\begin{align}
\iota_{X_{H^\star}}d\theta^\star =  -dH^\star.\label{dark_ham_equation}
\end{align}
We will refer to Eq.\,\eqref{dark_ham_equation} as the dark-plasma Hamilton's equations. We remark that the preceding discussion implies the formula $\dot{x} = X_{H^\star}(x)$ for dynamics on the slow manifold must be equivalent to the general formula for slow dynamics $\dot{x} = g_\epsilon(x,y^\star_\epsilon(x))$ to all orders.

At this point, we could in principle compute the Poisson tensor associated with the dark $2$-form $\Omega^\star = -d\theta^\star$ in order to express dark plasma dynamics in Poisson bracket form. However, such a computation would not be compatible with our desire to identify truncations of the formal power series $X_{H^\star}$ that possess a Hamiltonian structure. The essential issue is that the Poisson tensor associated with $\Omega^*$ is an infinite formal power series in $1/c$, naive truncations of which will fail to satisfy the Jacobi identity. This makes finding computable approximations of $X_{H^\star}$ with a true Hamiltonian structure extremely challenging. In the following subsection, we will therefore take a more nuanced approach to expressing dark plasma dynamics as a Hamiltonian system that makes structure-preserving truncation much simpler.

Our approach will follow the example set in \citet{burby2017magnetohydrodynamic}, where the same issue was addressed in the context of the slow manifold underlying magnetohydrodynamics. In particular, we will apply a near-identity non-canonical transformation $\mathcal{T}:x\mapsto \overline{x}$ on the dark slow manifold that causes the power series expansion of $\overline{\Omega}^\star = \mathcal{T}_*\Omega^\star$ to truncate at finite-order. (In fact we will achieve $\overline{\Omega}^\star = \overline{\Omega}^\star_0 + \tfrac{1}{c}\overline{\Omega}^\star_1$.) We say that the transformation $\mathcal{T}$ \emph{rectifies} the symplectic structure. This will allow us to identify a closed-form expression for the dark Poisson bracket at the cost of introducing some additional complexity into the dark Hamiltonian $\overline{H}^\star = \mathcal{T}_*H^\star$. Then we will compute the power series expansion of $\overline{H}^\star$ to the first post-Darwin order. By replacing $\overline{H}^\star$ with its post-Darwin approximation in the dark Hamilton equation $\iota_{\overline{X}_{\overline{H}^\star}}\overline{\Omega}^\star = d\overline{H}^\star$, while retaining the full (exact) form of $\overline{\Omega}^\star$, we will obtain a computable post-Darwin Hamiltonian approximation of dark plasma dynamics.

\subsection{Derivation of the rectifying transformation\label{Derivation of the rectifying transformation}}
The Lagrange $1$-form restricted to the dark slow manifold is given in Eq.\,\eqref{dadrk_Lagrange_form}. Note that the ``coordinates" we use on the dark slow manifold are $x = (g_\sigma,\mathbf{E}_L)$, where $g_\sigma$ is the species-$\sigma$ Lagrangian configuration map and $\mathbf{E}_L$ is the longitudinal eletric field. Since the constraint functions $\mathbf{A}_T^\star$ and $\mathbf{E}_T^\star$ that define the slow manifold are infinite formal power series in $1/c$, it is clear that $\theta^\star$ is also an infinite formal power series in $1/c$. We wish to change variables from $x$ to $\overline{x}$ using a near-identity transformation $\mathcal{T}:x\mapsto\overline{x}$ such that the power series defining the dark Lagrange $1$-form truncates modulo exact $1$-forms at finite order when expressed in terms of $\overline{x}=(\overline{g}_\sigma,\overline{\mathbf{E}}_L)$. We will call the transformation $\mathcal{T}$ a \emph{rectifying transformation}. We work modulo addition of exact $1$-forms to $\theta^\star$ because it is only the dark $2$-form $\Omega^\star = -d\theta^\star$ that is physically important.

It may not be immediately clear why it should be possible to find a rectifying transformation. The explanation dwells in the fact that symplectic manifolds have no local invariants. This should be contrasted with Riemannian geometry, where the curvature tensor is a local invariant; curvature cannot be eliminated by a coordinate transformation. Thus, symplectic manifolds satisfy the Darboux theorem, which says that, locally, all symplectic forms are related by coordinate transformations. While existence of a rectifying transformation, which we require to be globally defined, is not implied directly by the Darboux theorem, one can prove existence of such a transformation (as a formal power series) using the idea underlying Moser's celebrated proof of the Darboux theorem. See \citet{burby2017magnetohydrodynamic} for details.

In order to streamline our derivation, we represent $\mathcal{T}$ as a composition of formal Lie transforms, i.e. 
\begin{align}
\mathcal{T} = \dots\circ\exp(G_3)\circ \exp(G_2)\circ \exp(G_1),\label{Lie_transform_ansatz}
\end{align}
where the $G_k$ are vector fields on $x$-space (not to be confused with the $\mathbf{x}$-space $\mathbb{Q}$!) that we allow to be formal power series in $1/c$. To specify $\mathcal{T}$, we will derive formulas for the $G_k$. Our derivation will be facilitated by the well-known expression for the pushforward of a differential form $\alpha$ along a Lie transform $\exp(G)$:
\begin{align}
\exp(G)_*\alpha =\alpha -\mathcal{L}_{G}\alpha + \frac{1}{2}\mathcal{L}_{G}^2\alpha - \frac{1}{6}\mathcal{L}_G^3\alpha + \dots
\end{align}
In order to work with these infinite-dimensional Lie transforms explicitly, we will always compute Lie derivatives by first applying Cartan's identity $\mathcal{L}_{G} = \iota_{G}d + d\iota_G$ and then using the formalism described in Appendix \ref{inft_forms} for computing exterior derivatives on infinite-dimensional spaces.

The first step in our derivation is to list the leading five terms in the power series expansion of the (pre-transformed) dark Lagrange $1$-form $\theta^\star$. We have $\theta^\star =\theta^\star_0 + c^{-1}\,\theta^\star_1 + c^{-2}\theta^\star_2 + \dots$ with
\begin{align}
\iota_{\dot{x}}\theta^\star_{0\,x} & = \sum_\sigma \int \mathbf{p}\cdot V_\sigma^\mathbf{r}\,f_\sigma\,d^3r\,d^3p\label{theta_star_0}\\
\iota_{\dot{x}}\theta^\star_{1\,x} & = \sum_\sigma \int e_\sigma\mathbf{A}_L\cdot V_\sigma^{\mathbf{r}}\,f_\sigma\,d^3r\,d^3p - \frac{1}{4\pi}\int \mathbf{E}_L\cdot\dot{\mathbf{A}}_L\,d^3r\label{theta_star_1}\\
\iota_{\dot{x}}\theta^\star_{2\,x} & = \sum_\sigma \int e_\sigma \mathbf{A}_{T1}^\star\cdot V_\sigma^{\mathbf{r}}\,f_\sigma\,d^3r\,d^3p\label{theta_star_2}\\
\iota_{\dot{x}}\theta^\star_{3\,x}& =0\label{theta_star_3}\\
\iota_{\dot{x}}\theta^\star_{4\,x}& = \sum_\sigma \int e_\sigma \mathbf{A}_{T3}^\star\cdot V_\sigma^{\mathbf{r}}\,f_\sigma\,d^3r\,d^3p + \frac{1}{4\pi}\sum_\sigma\int \mathbf{E}_{T2}^\star\cdot D_{f_\sigma}\mathbf{A}_{T1}^\star[\text{div}(V_\sigma f_\sigma)]\,d^3r\label{theta_star_4}
\end{align}
where we have used $\mathbf{A}_{T0}^\star=\mathbf{A}_{T2}^\star = \mathbf{E}_{T0}^\star = \mathbf{E}_{T1}^\star =0$ and $D_{\mathbf{E}_L}\mathbf{A}_{T1}^\star = 0$. Higher-order contributions to $\theta^\star$ will not be necessary in our analysis. {{With the help of~\eqref{A1star}}  the second term of~\eqref{theta_star_4} can be written as }
\begin{equation}\label{lastoftheta_star_4}
  {  \sum_\sigma \frac{4\pi e_\sigma}{m_\sigma}\int f_\sigma\,\iota_{V_\sigma}d\left({\bf p}\cdot\Delta^{-1} \mathbf{E}_{T2}^\star\right) },
\end{equation}
where we have re-written 6-dimensional divergence as a Lie derivative and used the Cartan identity.
Next we record formulas for the exterior derivatives of $\theta^\star_k$ for $k=0,\dots,3$. (We will not need $d\theta^\star_4$.) In deriving these formulas, we reuse the procedure from Appendix \ref{inft_forms} for computing exterior derivatives in infinite dimensions.
\begin{align}
\iota_{\delta x_2}\iota_{\delta x_1}d\theta^\star_{0\,x}& = \sum_\sigma\int \iota_{\xi_{\sigma 2}}\iota_{\xi_{\sigma 1}}d\theta_0\,f_\sigma\,d^3r\,d^3p\label{dtheta_0}\\
\iota_{\delta x_2}\iota_{\delta x_1}d\theta^\star_{1\,x}& =\sum_\sigma\int e_\sigma\left(\delta\mathbf{A}_{L1}\cdot\xi_{\sigma 2}^{\mathbf{r}} - \delta\mathbf{A}_{L2}\cdot\xi_{\sigma 1}^{\mathbf{r}}\right)f_\sigma\,d^3r\,d^3p\nonumber\\
& - \frac{1}{4\pi}\int\left(\delta\mathbf{E}_{L1}\cdot\delta\mathbf{A}_{L2} - \delta\mathbf{E}_{L2}\cdot\delta\mathbf{A}_{L1}\right)\,d^3r\label{dtheta_1}\\
\iota_{\delta x_2}\iota_{\delta x_1}d\theta^\star_{2\,x}& =\sum_\sigma \int e_\sigma \mathbf{B}_1^\star\cdot \xi_{\sigma 1}^{\mathbf{r}}\times\xi_{\sigma 2}^{\mathbf{r}}\,f_\sigma\,d^3r\,d^3p\nonumber\\
&\hspace{-7em}+4\pi\int \left(\sum_\sigma\frac{e_\sigma}{m_\sigma}\nabla\cdot\int \xi_{\sigma 1}^{\mathbf{r}}\mathbf{p}\,f_\sigma\,d^3p\right)\cdot\Delta^{-1}\Pi_T\left(\sum_\sigma e_\sigma\int \xi_{\sigma 2}^{\mathbf{r}}\,f_\sigma\,d^3p\right)\,d^3r\nonumber\\
&\hspace{-7em}-4\pi\int \left(\sum_\sigma\frac{e_\sigma}{m_\sigma}\nabla\cdot\int \xi_{\sigma 2}^{\mathbf{r}}\mathbf{p}\,f_\sigma\,d^3p\right)\cdot\Delta^{-1}\Pi_T\left(\sum_\sigma e_\sigma\int \xi_{\sigma 1}^{\mathbf{r}}\,f_\sigma\,d^3p\right)\,d^3r\nonumber\\
&\hspace{-7em}-4\pi\int \left(\sum_\sigma\frac{e_\sigma}{m_\sigma}\int \xi_{\sigma 1}^{\mathbf{p}}\,f_\sigma\,d^3p\right)\cdot\Delta^{-1}\Pi_T\left(\sum_\sigma e_\sigma\int \xi_{\sigma 2}^{\mathbf{r}}\,f_\sigma\,d^3p\right)\,d^3r\nonumber\\
&\hspace{-7em}+4\pi\int \left(\sum_\sigma\frac{e_\sigma}{m_\sigma}\int \xi_{\sigma 2}^{\mathbf{p}}\,f_\sigma\,d^3p\right)\cdot\Delta^{-1}\Pi_T\left(\sum_\sigma e_\sigma\int \xi_{\sigma 1}^{\mathbf{r}}\,f_\sigma\,d^3p\right)\,d^3r\label{dtheta_2}\\
\iota_{\delta x_2}\iota_{\delta x_1}d\theta^\star_{3\,x}& =0. \label{dtheta_3}
\end{align}
We now determine $G_1$

by requiring it solves the linear equation
\begin{align}
\iota_{G_1}d\theta^\star_0 +c^{-1}\iota_{G_1}d\theta^\star_1 = c^{-2}\theta^\star_2.\label{G1_homological_equation}
\end{align}
The steps necessary in inverting this equation have be outlined in equations~\eqref{VlasovPoissonVF},~\eqref{VdotH},~\eqref{EdotH} and~\eqref{AdotH}. (The reason for this choice of $G_1$ will become clear soon.) Using the formulas \eqref{dtheta_0}, \eqref{dtheta_1}, and \eqref{theta_star_2}, we find that $G_1 = (Y_{\sigma1}\circ g_\sigma, G_1^{\mathbf{E}_L},G_1^{\mathbf{A}_L})$ is given by
\begin{align}
Y_{\sigma1} & = \frac{e_\sigma}{c^2}\mathbf{A}_{T1}^\star\cdot\partial_{\mathbf{p}}\label{Y1_formula}\\
G_1^{\mathbf{E}_L} & = 0\label{G1EL_formulua}\\
G_1^{\mathbf{A}_L} & = 0.
\end{align}
Since $G_1$ is $O(c^{-2})$ and satisfies Eq.\,\eqref{G1_homological_equation}, the pushforward of $\theta^\star$ along the Lie transform $\exp(G_1)$ has the power series expansion
\begin{align}
\exp(G_1)_*\theta^\star & = \theta^\star_0 + c^{-1}\theta^\star_1 + c^{-4}\theta^\star_4 - \frac{1}{2}c^{-2}\,\iota_{G_1}d\theta^\star_2+ O(c^{-5}),
\end{align}
where we have used our freedom to add exact differentials to the Lagrange $1$-form in order to make the replacement $\exp(G_1)_* = \exp(-\mathcal{L}_{G_1})\rightarrow \exp(-\iota_{G_1}d)$.
Thus, if we were to stop here and set $G_k = 0$ for $k>1$ we would have succeeded in making the difference between the transformed Lagrange $1$-form $\overline{\theta}^\star$ and $\theta^\star_0 + c^{-1}\theta^\star_1$ one order higher than prior to applying the transformation.

Finally, we turn to determining the $G_k$ for $k>1$. Since the first Lie transform, generated by $G_1$, brought the Lagrange $1$-form closer to $\theta^\star_0 + c^{-1}\,\theta^\star_1$, we will select the higher $G_k$ so that the transformed Lagrange form is equal to $\theta^\star_0 + c^{-1}\,\theta^\star_1$ to all orders in $1/c$. We give an inductive proof in Appendix \ref{rect_exist_sec} that this can indeed be done. For the purposes of this Article, it is only necessary for us to explicitly perform the algebra that determines $G_2$. However, all subsequent $G_k$ may be determined in a similar manner. (The tedium of the required algebraic manipulations quickly becomes unreasonable for $k>2$.)

In keeping with the above strategy, we demand that $G_2$ satisfies the linear equation
\begin{align}
\iota_{G_2}d\theta^\star_0+c^{-1}\iota_{G_2}d\theta^\star_1 = c^{-4}\theta^\star_4 - \frac{1}{2}c^{-2}\iota_{G_1}d\theta_2^\star.\label{g2_algebraic_eqn}
\end{align}
{What remain is to estimate the last term.} {From the last four formidable-looking lines of~\eqref{dtheta_2}, only the third one will contribute to the final result because of the form of $G_1$ as we see from the equation~\eqref{g2_algebraic_eqn}.}
Using formulas \eqref{dtheta_0}, \eqref{dtheta_1}, \eqref{theta_star_4},~\eqref{lastoftheta_star_4} and \eqref{dtheta_2}, we find the solution $G_2 = (Y_{\sigma2}\circ g_\sigma,G_2^{\mathbf{E}_L},G_2^{\mathbf{A}_L})$ is given by
\begin{align}
Y_{\sigma 2} &=\frac{e_\sigma}{c^4}\left[\mathbf{A}_{T3}^\star + \nabla\left(\frac{\mathbf{p}}{m_\sigma}\cdot\Delta^{-1}\mathbf{E}_{T2}^\star\right)+\frac{1}{2}\Delta^{-1}\Pi_T\omega_p^2\mathbf{A}_{T1}^\star\right]\cdot\frac{\partial}{\partial\mathbf{p}}\nonumber\\
&-\frac{1}{c^4}\left[\frac{e_\sigma}{m_\sigma}\Delta^{-1}\mathbf{E}_{T2}^\star\right]\cdot\frac{\partial}{\partial\mathbf{r}}\\
G_2^{\mathbf{E}_L} & = \frac{1}{c^4}\Pi_L\omega_p^2\Delta^{-1}\mathbf{E}_{T2}^\star\\
G_2^{\mathbf{A}_L} & = 0.
\end{align}
These formulas show in particular that $G_2 = O(c^{-4})$. Since it can be shown that $G_k = O(c^{-5})$ for $k> 2$, this, together with Eq.\, \eqref{g2_algebraic_eqn} implies 
\begin{align}
\overline{\theta}^\star = \theta^\star_0 + c^{-1}\,\theta^\star_1 + O(c^{-5}),
\end{align}
consistent with our claim that $\overline{\theta}^\star = \theta^\star_0 + c^{-1}\,\theta^\star_1$ to all orders in $1/c$. 

To summarize, we have identified a near-identity non-canonical transformation on the dark slow manifold that rectifies the dark Poisson bracket. Using the variables $(g_\sigma, \mathbf{E}_L,\mathbf{A}_L)$ to parameterize the slow manifold, the transformation is given explicitly to $O(c^{-4})$ by 
\begin{align}
{\mathbf{A}}_L & =\overline{\mathbf{A}}_L\label{inverse_one}\\
{\mathbf{E}}_L & =\overline{\mathbf{E}}_L - \sum_\sigma\f{1}{c^4} \Pi_L \overline{\omega}_{p\sigma}^2\Delta^{-1} \overline{{\bf E}}_{T2}^\star + O(c^{-5})\\
{{f}}_\sigma & =\overline{{f}}_\sigma + \frac{e_\sigma}{c^2}\overline{\mathbf{A}}_{T1}^\star\cdot \partial_{\mathbf{p}}\overline{{f}}_\sigma +\frac{1}{2}\frac{e_\sigma^2}{c^4}\overline{\mathbf{A}}_{T1}^*\overline{\mathbf{A}}_{T1}^*:\partial_{\mathbf{p}}^2\overline{{f}}_\sigma\nonumber\\ 
&+\f{e_\sigma}{c^4}\lf[\overline{{\bf A}}_{T3}^\star + \f{1}{2} \Delta^{-1}{\Pi_T}\overline{\omega}_{p}^2 \overline{{\bf A}}_{T1}^\star  + \Delta^{-1} \nabla \f{{\bf p}\cdot \overline{{\bf E}}_{T2}^\star}{m_\sigma} \ri]\cdot \partial_{\mathbf{p}}\overline{{f}}_\sigma \nonumber\\
&- \f{e_\sigma}{m_\sigma c^4} \Delta^{-1} \overline{{\bf E}}^\star_{T2}\cdot\nabla \overline{f}_\sigma+O(1/c^5).\label{inverse_three}
\end{align}
These formulas follow from the definition of a Lie transform; if $G = (Y_\sigma\circ g_\sigma,G^{\mathbf{E}_L},G^{\mathbf{A}_L})$ is the Lie generator then $\exp(G)(g_\sigma,\mathbf{E}_L,\mathbf{A}_L) = (\overline{g}_\sigma,\overline{\mathbf{E}}_L,\overline{\mathbf{A}}_L)$ is the $\lambda = 1$ solution of the system of evolution equations
\begin{align}
\frac{dg_\sigma}{d\lambda} = Y_\sigma\circ g_\sigma,\quad \frac{d\mathbf{E}_L}{d\lambda} = G^{\mathbf{E}_L},\quad \frac{d\mathbf{A}_L}{d\lambda} = G^{\mathbf{A}_L}.
\end{align}
After applying this rectifying transformation, the dark Lagrange $1$-form is given by $\overline{\theta}^\star = \theta^\star_0 + c^{-1}\,\theta^\star_1$ to all orders in $1/c$. Therefore the (rectified) dark symplectic form $-d\overline{\theta}^\star =\overline{\Omega}^\star$is given by
\begin{align}
{\iota_{\delta x_2}\iota_{\delta x_1}}\overline{\Omega}^\star
& = \sum_\sigma\int \omega_0[\xi_{\sigma 1},\xi_{\sigma 2}]\,f_\sigma\,d^3r\,d^3p + \frac{1}{4\pi c}\int (\delta\mathbf{E}_{L1}\cdot\delta\mathbf{A}_{L2} - \delta\mathbf{E}_{L2}\cdot\delta\mathbf{A}_{L1})\,d^3r\nonumber\\
&-\sum_\sigma\int \frac{e_\sigma}{c}(\delta \mathbf{A}_{L1}\cdot\xi_{\sigma 2}^{\mathbf{r}} - \delta\mathbf{A}_{L2}\cdot\xi_{\sigma 1}^{\mathbf{r}})\,f_\sigma\,d^3r\,d^3p,
\end{align}
and the associated dark Poisson bracket is
\begin{eqnarray}\label{vlasovpoissonbracket}
\{\calF,\calG\}^\star &=& \sum_\sigma\int d^6 z \, f_\sigma \lf[\delf{\calF}{g_\sigma}\circ g^{-1}_\sigma -4\pi e \delf{\calF}{{\bf E}_L}\cdot d\mathbf{r}\ri]\cdot \mathcal{J}_0 \cdot \lf[\delf{\calG}{g_\sigma}\circ g^{-1}_\sigma  -4\pi e \delf{\calG}{{\bf E}_L}\cdot d\mathbf{r}\ri]\nonumber \\
&+& 4\pi c\int d^3r \lf(\delf{\calF}{{\bf E}_L}\cdot \delf{\calG}{{\bf A}_L} -  \delf{\calG}{{\bf E}_L}\cdot \delf{\calF}{{\bf A}_L} \ri).
\end{eqnarray}
We remark that the functional derivative with respect to a longitudinal vector field is defined (consistently and in accordance with the standard mathematical notion of functional derivative) to be a longitudinal vector field. Therefore in particular if $\mathbf{W}$ is some vector field then $\delta/\delta\mathbf{E}_L\int \mathbf{E}_L\cdot \mathbf{W}\,d^3 r = \Pi_L\mathbf{W}$, \emph{not} $\mathbf{W}$. We also remark that the bracket $\{\cdot,\cdot\}^\star$ is written in terms of Lagrangian configuration maps; the corresponding Eulerian bracket is
\begin{eqnarray}\label{darkbracket}
\{\calF,\calG\}_{\text{Dark}} &=& \sum_\sigma\int d^6 z \, f_\sigma \lf[d\delf{\calF}{f_\sigma} -4\pi e \delf{\calF}{{\bf E}_L}\cdot d\mathbf{r}\ri]\cdot \mathcal{J}_0 \cdot \lf[d\delf{\calG}{f_\sigma}  -4\pi e \delf{\calG}{{\bf E}_L}\cdot d\mathbf{r}\ri]\nonumber \\
&+& 4\pi c\int d^3r \lf(\delf{\calF}{{\bf E}_L}\cdot \delf{\calG}{{\bf A}_L} -  \delf{\calG}{{\bf E}_L}\cdot \delf{\calF}{{\bf A}_L} \ri).
\end{eqnarray}


\subsection{Derivation of the post-Darwin Hamiltonian\label{Calculation of the dark Poisson bracket and post-Darwin Hamiltonian}}

In the previous subsection we succeeded in identifying a rectifying transformation that allowed us to find a closed-form expression for the Poisson bracket on the slow manifold. We are now in position to derive approximations of the dark plasma evolution equations that possess a Hamiltonian structure. The remaining step is to derive a formula for the slow manifold Hamiltonian expressed in terms of the rectified variables $(\overline{g}_\sigma,\overline{\mathbf{E}}_L,\overline{\mathbf{A}}_L)$. The purpose of this section is to derive that Hamiltonian $\overline{H}^\star$ to the first post-Darwin order. Our post-Darwin approximation to dark plasma dynamics is then defined by the Hamilton equation $\dot{x} = \{x,\overline{H}^\star_{PD}\}_{\text{Dark}}$, where $\overline{H}^\star_{PD}$ is the post-Dawin Hamiltonian and the rectified dark Poisson bracket is given in Eq.\,\eqref{vlasovpoissonbracket}.

The original Hamiltonian~\eqref{MVHam} reads
\begin{equation}
H^\star(f_\sigma,{\bf E}_L) = \sum_\sigma\int d^6 z\, f_\sigma \gamma_\sigma m_\sigma c^2 + \f{1}{8\pi} \int d^3r \lf(  E^2_L  + E^{\star 2}_T +  B^{\star 2} \ri)
\end{equation}
and can be expended in powers of $1/c$ as $H^\star(f,E_L) = c^2 H_{-2}^\star + H_0^\star + c^{-2} H_2^\star + c^{-4} H_2$ with the two lowest order terms given as
\begin{equation}\label{H0order}
H^\star_{-2} = \sum_\sigma \int d^6 z\, f_\sigma m_\sigma c^2,\quad H^\star_{0} =\sum_\sigma \int d^6 z\, \f{f_\sigma p^2}{2m_\sigma} + \f{1}{8\pi}\int d^3r\, E_L^2,
\end{equation}
the (quadratic in $f_\sigma$) Darwin terms given as
\begin{equation}\label{H2order1}
H^\star_2 = - \sum_\sigma\int d^6 z \, \f{f_\sigma p^4}{8 m^3_\sigma} + \f{1}{8\pi}\int d^3r \, { B}^{\star 2}_{1},
\end{equation}
and the ``piezoelectric terms" given as
\begin{equation}\label{H4order}
H^\star_4 = \sum_\sigma\int d^6 z \f{f_\sigma p^6}{16 m^5_\sigma} + \f{1}{8\pi} \int d^3 r\, |{\bf E}_{T2}^\star|^{ 2} + \f{1}{4\pi} \int d^3 r\, {\bf B}_{1}^\star\cdot {\bf B}_{3}^\star.
\end{equation}
Note that the second term in $H^\star_4$ is quadratic in $f_{\sigma}$, while the last term is cubic. The cubic term, which involves $\mathbf{B}^\star_3$ appears particularly formidable. (C.f. Eq.\,~\eqref{B3star}). As we will soon see, much of the complexity introduced by this term is eliminated by the rectifying transformation~\eqref{inverse_three}.
Also note that we are free to remove the rest-energy term $H^\star_{-2}$ since it is a Casimir. 

By defining  $\boldsymbol{ \pi}_\sigma:= {\bf p} - \frac{e_\sigma}{c^2} {\bf A}_{T1}^\star$ and substituting the inverse transformation formulas \eqref{inverse_one}-\eqref{inverse_three} into the Hamiltonian on the slow manifold, we obtain the following expression for {the post-Darwin Hamiltonian} in terms of the rectified variables $(\overline{\mathbf{E}}_L,\overline{\mathbf{A}}_L,\overline{f}_\sigma)$
\begin{align}
&{\overline{H}^\star_{\text{PD}}}(\overline{\mathbf{E}}_L,\overline{f}_\sigma) =\sum_\sigma \int \left(\frac{|\overline{\boldsymbol{\pi}}_\sigma|^2}{2m_\sigma} - \frac{|\overline{\boldsymbol{\pi}}_\sigma|^4}{8 m_\sigma^3 c^2}+ \frac{|\overline{\boldsymbol{\pi}}_\sigma|^6}{16 m_\sigma^5c^4}\right)\,\overline{f}_\sigma\,d^3r\,d^3p\nonumber\\
-&\frac{2\pi}{c^2}\int \left[\Pi_T\overline{\mathbf{J}}_0 + \Pi_T\tfrac{\overline{\omega}_p^2}{c^2}\Delta^{-1}\Pi_T\overline{\mathbf{J}}_0\right]\Delta^{-1}\left[\Pi_T\overline{\mathbf{J}}_0 + \Pi_T\tfrac{\overline{\omega}_p^2}{c^2}\Delta^{-1}\Pi_T\overline{\mathbf{J}}_0\right]\,d^3r\nonumber\\
{-}&\frac{{ 1}}{8\pi c^4}\int |\overline{\mathbf{E}}_{T2}^\star|^2\,d^3r +\frac{1}{8\pi}\int |\overline{\mathbf{E}}_L|^2\,d^3r +\frac{2\pi}{ c^4}\int \overline{\omega}_p^2\,|\Delta^{-1}\Pi_T\overline{\mathbf{J}}_0|^2\,d^3r,\label{dark_ham}
\end{align}
where special care was made to write the expression so that it is manifestly positive-definite. The integrand on the first line comprises the first three non-constant terms of the Taylor expansion of $\sqrt{1+|\overline{\boldsymbol{\pi}}_\sigma|^2/m_\sigma^2 c^2}$, which is readily seen to be positive. The second line involves a $\mathcal{O}(c^{-6})$ term which will be dropped below to obtain the $1/c$ expansion of the equations of motion, but otherwise it is quartic in $f$.
Modulo $O(c^{-6})$ terms, the second line and the last term in the third line can be condensed into
\begin{equation}
    -\frac{1}{8\pi c^2} \int d^3 r\, \overline{\bf A}_{T1}^\star \cdot \left(\Delta + \frac{\overline{\omega}_p^2}{c^2} \right) \overline{\bf A}_{T1}^\star.
\end{equation}
 If we combine everything together and take variational derivatives we get
\begin{eqnarray}
    \frac{\delta \overline{H}^\star_{PD}}{\delta \overline{f}_\sigma} =
    {\frac{p^2}{2m_\sigma}-\frac{p^4}{8m_\sigma^3 c^2}+\frac{p^6}{16m_\sigma^5c^4}}
    {-\frac{e_\sigma {\bf p}\cdot {\bf A}^\star_{T1}}{m_\sigma c^2}\lf(1-\frac{p^2}{2m^2_\sigma c^2} \ri)}
     \nonumber\\+{ \frac{4\pi e_\sigma}{m_\sigma c^4} {\bf p}\cdot \Delta^{-1}\Pi_T J_2 }
    {-}\frac{{ e_\sigma} }{c^4}\left( \frac{\bf p p}{{m_\sigma}^2} : \Delta^{-1}\nabla \,\overline{\bf E}_{T2}^\star + \frac{e_\sigma}{m_\sigma} \overline{\bf E}_L\cdot \Delta^{-1} \overline{\bf E}_{T2}^\star \right),
\end{eqnarray}
where the square roots indicate the expansion introduced in~\eqref{dark_ham} and we have used equation~\eqref{jexpansion}, while the variational derivative with respect to the other remaining slow variable reads
\begin{equation}
    \frac{\delta \overline{H}^\star_{PD}}{\delta \overline{\bf E}_L} = \frac{\overline{ \bf E}_L}{4\pi} - \frac{{1}}{4\pi c^4} {\Pi_L}\overline{\omega}_p^2 \Delta^{-1} \overline{\bf E}_{T2}^\star
\end{equation}
This completes the derivation of the post-Darwin Hamiltonian. 

Using the Poisson bracket \eqref{darkbracket}, the corresponding equations of motion on the slow manifold are
\begin{gather}
\partial_t \overline{f}_\sigma + \nabla\cdot\left(\left[\partial_{\mathbf{p}}\frac{\delta \overline{H}^\star_{PD}}{\delta f_\sigma}\right]\overline{f}_\sigma\right) + \partial_{\mathbf{p}}\cdot\left(\left[4\pi e_\sigma \frac{\delta \overline{H}^\star_{PD}}{\delta\mathbf{E}_L} - \nabla\frac{\delta \overline{H}^\star_{PD}}{\delta f_\sigma}\right]\overline{f}_\sigma\right) = 0\label{dark_hamiltonian_vlasov}\\
\partial_t \overline{\mathbf{E}}_L +4\pi\Pi_L\sum_\sigma e_\sigma \int \left(\partial_{\mathbf{p}}\frac{\delta \overline{H}^\star_{PD}}{\delta \overline{f}_\sigma}\right)\,\overline{f}_\sigma\,d^3p = 4\pi c\frac{\delta \overline{H}^\star_{PD}}{\delta \overline{\mathbf{A}_L}} \label{longitudinal_ampere}\\
\partial_t\overline{\mathbf{A}}_L = -4\pi c\frac{\delta \overline{H}^\star_{PD}}{\delta \overline{\mathbf{E}}_L}.\label{longitudinal_vector_potential}
\end{gather}
Note that since $\overline{H}^\star_{PD}$ is independent of $\mathbf{A}_L$ Eq.\,\eqref{longitudinal_ampere} reproduces the longitudinal Amp\`ere equation, which is equivalent to the preservation of Gauss's Law. This result will persist to all orders in $c^{-1}$ as a result of the gauge invariance of the rectifying transformation. For the same reason, Eqs.\,\eqref{dark_hamiltonian_vlasov} and \eqref{longitudinal_ampere} comprise a closed system of evolution equations for the distribution function $\overline{f}_\sigma$ and the longitudinal electric field $\overline{\mathbf{E}}_L$. As such, the evolution equations for dark 
{manifold} dynamics may be written in terms of the same dependent variables used in the Vlasov-Poisson system.


\subsection{On the Hamiltonian nature of Darwin's approximation}
We can collect all terms up to $1/c^2$ obtained in the previous section to get the transformed Darwin's Hamiltonian
\begin{eqnarray}
    \overline{H}_D^\star(\overline{f}_\sigma,\overline{\bf E}_L) &=& \sum_\sigma \int d^6 z\, \f{\overline{f}_\sigma p^2}{2m_\sigma} + \f{1}{8\pi}\int d^3r\, \overline{E}_L^2 \nonumber\\
    &-&\sum_\sigma\int d^6 z \, \f{\overline{f}_\sigma p^4}{8 m^3_\sigma c^2} +\f{2\pi}{c^2}\int d^3r\, \overline{\bf J}_{0} \cdot  \Delta^{-1} \Pi_T\, \overline{\bf J}_{0}, 
\end{eqnarray}
We note here that, to this order, the transformations~\eqref{inverse_three} amount to a transformation from kinetic to canonical momentum.

While the Hamiltonian approach to post-Darwin's extension appears to be new, there is an instance of a Hamiltonian study of Darwin's approximation found in literature, namely \citet{krause07}. In particular, in this work action principles for the Darwin approximation in the Vlasov context were presented. However, the phase space only spans the field of density functions $f$, so that the Lie-Poisson structure is that of a Vlasov-Poisson bracket
\begin{equation}\label{krausebracket}
    \{F, G\}_{D}=\int d^{6} \zeta f_{D}\left[\frac{\delta F}{\delta f_{D}}, \frac{\delta G}{\delta f_{D}}\right]
\end{equation}
which naturally produces Vlasov equation ${\partial f_{D}}/{\partial t}=\left\{f_{D}, H\right\}$ when equipped with the Hamiltonian
\begin{equation}\begin{aligned}
H\left[f_{D}\right]=& \int d^{6} \zeta f_{D}(\zeta, t)\left[\frac{\pi^{2}}{2 m}-\frac{\pi^{4}}{8 m^{3} c^{2}}\right] \\
&+\frac{e^{2}}{2} \int d^{6} \zeta \int d^{6} \zeta^{\prime} f_{D}(\zeta, t) f_{D}\left(\zeta^{\prime}, t\right) K\left(\mathbf{r} | \mathbf{r}^{\prime}\right) \\
&-\frac{e^{2}}{2 m^{2} c^{2}} \int d^{6} \zeta \int d^{6} \zeta^{\prime} f_{D}(\zeta, t) f_{D}\left(\zeta^{\prime}, t\right) K_{i j}\left(\mathbf{r} | \mathbf{r}^{\prime}\right) \pi_{i} \pi_{j}^{\prime},
\end{aligned}\end{equation}
where $\pi_i$ correspond to eulerianized full canonical momenta. To make a further comparison we recognize from eqs. (15), (16) and (23) of \citet{krause07} that $\int d^3 r^\prime \, K_{ij}\left(\mathbf{r} | \mathbf{r}^{\prime}\right) \equiv -4\pi\Delta^{-1}\Pi_T$ and that they have single species. In addition, in our case (eq.~\eqref{dark_ham}), $\boldsymbol{ \pi}$ is a kinetic momentum up to $\mathcal{O}(1/c^2)$, since the active transformations we performed in~\eqref{inverse_three} can be viewed as a passive relabeling transformation in ${\bf p}$. So the $\pi_i$ in~\citet{krause07} correspond to our new ${\bf p}$ and not $\boldsymbol{\pi}$

It is straightforward to recover this Hamiltonian formulation of the Darwin equation from ours by restricting to a level set of the momentum map associated with gauge symmetry (i.e. the residual of Gauss's Law) and then quotienting by translations in $\mathbf{A}_L$.


\section{Conclusion}
In this article we applied slow manifold reduction to a particular non-relativistic scaling of the Maxwell--Vlasov system. The main motivation was 
{to develop a Hamiltonian post-Darwin approximation. In the process we obtained the Braginskii pressure tensor in the second order and curiously static magnetic field response to the heat flux tensor. }

Another, and perhaps more important application of these Hamiltonian asymptotics concerns numerical calculations. Previously, various numerical integrators have benefited from Darwin's approximation in convergence of their algorithms. For instance, \citet{CHEN201573} introduce a  conservative, nonlinearly implicit PIC algorithm for the Vlasov–Darwin system. The motivation was avoiding spurious radiative noise present in fully implicit, energy conserving Maxwell--Vlasov implementations, when employing large implicit timesteps for multiscale, lowfrequency problems. The second order expansion that we obtained can be applied in the same spirit as Darwin's approximation in order to reduce the integration time of Vlasov codes/ (particle-in-cell) PIC simulations but with greater fidelity. 

The Poisson bracket describing dark plasma dynamics is defined \emph{a priori} as an infinite formal power series. This is problematic from a practical point of view since truncating the series will violate the Jacobi identity in general. Following \citet{burby2017magnetohydrodynamic}, we overcame this difficulty by applying a non-canonical near-identity transformation to the dark slow manifold that caused the transformed slow-manifold Poisson bracket to truncate exactly at finite order. Conveniently, this transformed bracket agreed with a well-known bracket for the Vlasov-Poisson system. An alternative approach to achieving the same result would be to apply \emph{canonical} near-identity transformations to the full Maxwell--Vlasov phase space as in \citet{Brizard_Chandre_2020} in order to make the dark slow manifold truncate at finite order. A benefit of this alternative approach is that it would facilitate the analysis of Maxwell--Vlasov dynamics on \emph{and near} the slow manifold, i.e. ``dim" plasma dynamics. We plan to pursue this idea in future work.

There are further avenues to extend this line of research. As mentioned above one can attempt to construct structure preserving algorithms for PIC simulations which reduce the integration times. This would thus require discretization techniques, which can be performed directly within the Hamiltonian action principle formalism. In the infinite dimensional case, as usual, with the noncanonical Hamiltonian formulation, energy-Casimir method comes to mind, which permits comprehensive study of the stability. In addition, one can envision the utility of the integral transform methods to the Darwinian and Piezoelectric approximations. Integral transforms, such as \emph{G}-transform were applied earlier to simplify the dynamics of Vlasov-Poisson in case of collisions present in the system by \citet{heninger18}.



\section*{Acknowledgements}

We would like to thank the hospitality of the Mathematical Sciences Research Institute in Berkeley during the semester of ``Hamiltonian systems, from topology to applications through analysis in 2018'', where much of this work was done. 

\section*{Funding}

This work was supported by the Los Alamos National Laboratory LDRD program under project 20180756PRD4.

\section*{Declaration of interest}

The authors report no conflict of interest.


\appendix



\section{Differential forms on infinite-dimensional manifolds\label{inft_forms}}
On finite-dimensional spaces, computations with differential forms may be reduced to repeated applications of the identity $d(f\,dg) = df\wedge dg$, for scalar functions $f,g$. For example, if $\alpha$ is a $1$-form then its exterior derivative may be computed by first introducing coordinates $x^i$, framing the cotangent bundle using the coordinate differentials $dx^i$, and then applying the aformentioned identity according to $d\alpha = d(\alpha_idx^i) = d\alpha_i\wedge dx^i = \partial_j\alpha_i\,dx^j\wedge dx^i$, where $\alpha_i$ denote the component functions of $\alpha$ in the basis $dx^i$. Lie derivatives $\mathcal{L}_X$ along a vector field $X$ may also be computed using this rule by first applying the Cartan formula $\mathcal{L}_X\lambda = \iota_Xd\lambda + d(\iota_X\lambda)$, where $\lambda$ is any differential form.

On infinite-dimensional spaces, however, the identity $d(f\,dg) = df\wedge dg$ is less useful for computing with forms. The essential issue is that the introduction of coordinates $x^i$ is, at best, more challenging in infinite-dimensions. On linear function spaces, one might occasionally employ Fourier coefficients, or some other well-known basis coefficients  successfully as coordinates. But constructing coordinates on nonlinear spaces such as the group of diffeomorphisms $g$ of a fixed manifold $M$ is much more involved. It would therefore be useful to develop a coordinate-independent formalism for computing with differential forms. Such a formalism would apply uniformly across finite-dimensional and infinite-dimensional spaces, at least at a formal level. The purpose of this appendix is to supply one such formalism, which is applied elsewhere in this Article.

For the purposes of this Article, it is sufficient to discuss only scalars ($0$-forms) $F$, $1$-forms $\alpha$, and the corresponding exterior derivatives $dF,d\alpha$. We assume these objects are defined on a manifold $M$ with points $m$, tangent spaces $T_mM$, and tangent vectors $\delta m\in T_mM$.
\\ \\
\noindent\emph{Exterior derviative of a $0$-form:}
The exterior derivative of a $0$-form $F$ is a $1$-form $dF$. Thus, $dF$ assigns a linear functional $dF_m:T_mM\rightarrow\mathbb{R}$ to each point $m\in M$. If $\delta m\in T_mM$ is a tangent vector at $m$ we denote the value of $dF_m$ applied to $\delta m$ as $\iota_{\delta m}dF_m\in \mathbb{R}$. To define $dF$, it is sufficient to specify the value of $\iota_{\delta m}dF_m$ for arbitrary $m$ and $\delta m$. To that end, we choose a curve $c(t)\in M$ such that $c(0)=m$ and $c^\prime(0) = \delta m$, where $c^\prime(t)$ denotes the velocity of $c(t)$. The value of $\iota_{\delta m}dF_m$ is then given by
\begin{align}\label{eq:idmDFm}
    \iota_{\delta m}dF_m {:=} \frac{d}{dt}\bigg|_0F(c(t)).
\end{align}
\\ \\
\noindent \emph{Exterior derivative of a $1$-form:}
The exterior derivative of a $1$-form $\alpha$ is a $2$-form $d\alpha$. Thus, $d\alpha$ assigns a skew-symmetric bilinear functional $d\alpha_m:T_mM\times T_mM\rightarrow\mathbb{R}$ to each point $m\in M$. If $\delta m_1,\delta m_2$ are two tangent vectors at $m$, we denote the value of $d\alpha_m$ applied to the pair $(\delta m_1,\delta m_2)$ as $\iota_{\delta m_2}\iota_{\delta m_1}d\alpha_m$. Note that skew-symmetry implies $\iota_{\delta m_2}\iota_{\delta m_1}d\alpha_m =- \iota_{\delta m_1}\iota_{\delta m_2}d\alpha_m$. To define $d\alpha_m$, it is sufficient to specify the value of $\iota_{\delta m_2}\iota_{\delta m_1}d\alpha_m$ for arbitrary $m$ and $(\delta m_1,\delta m_2)$. To that end, we introduce the functional $I$ on the space of curves $c(t)\in M$ with $c(-1)$ and $c(1)$ fixed whose value at $c$ is $I(c) {:=} \int_{-1}^{1}\iota_{c^\prime(t)}\alpha_{c(t)}\,dt$. We also introduce a {given} curve $c(t)\in M$ with $c(0)=m$ and $c^\prime(0)=\delta m_2$, and a vector field $\delta c(t)$ along $c(t)$ with $\delta c(0) = \delta m_1$, $\delta c(-1) =0$, and $\delta c(1) = 0$. Then we define
\begin{align}
    \iota_{\delta m_2}\iota_{\delta m_1}d\alpha_m {:=} \lim_{a\rightarrow 0}\frac{1}{2a} \delta I(c)[\mathbb{I}_{[-a,a]}\delta c],\label{d_one_form}
\end{align}
where $\delta I(c)[\delta c]$ denotes the first variation of $I$ at $c$ in the direction $\delta c$, and $\mathbb{I}_{[-a,a]}:[-1,1]\rightarrow\mathbb{R}$ denotes the indicator function for the interval $[-a,a]$. 

In finite dimensions, the formula \eqref{d_one_form} recovers the usual definition of exterior derivative because
\begin{align*}
     \lim_{a\rightarrow 0}\frac{1}{2a} \delta I(c)[\mathbb{I}_{[-a,a]}\delta c]&= \lim_{a\rightarrow 0}\frac{1}{2a}\int_{-a}^{a}\iota_{c^\prime(t)}\iota_{\delta c(t)}d\alpha_{c(t)}\,dt\\
    & = \iota_{c^\prime(0)}\iota_{\delta c(0)}d\alpha_{c(0)}\\
    & = \iota_{\delta m_2}\iota_{\delta m_1}d\alpha_{m}
\end{align*}
where we have used the indicator function to change the limits of integration and the Lebesgue differentiation theorem to evaluate the limit $a\rightarrow 0$. In infinite dimensions, \eqref{d_one_form} is useful because it does not require parameterizing $M$ with any linear space, as the following example illustrates.
\\ \\
\noindent \emph{Example on the diffeomorphism group}
We will illustrate how the formula \eqref{d_one_form} may be used to compute exterior derivatives when working on the diffeomorphism group of a fixed manifold $P$. 

If $g:P\rightarrow P$ is a diffeomorphism then a tangent vector at $g$ is a map $\delta g$ that assigns to each $p\in P$ a tangent vector at $g(p)$, i.e. $\delta g(p)\in T_{g(p)}P$. Note that if $\delta g$ is a tangent vector at $g$, then $\xi = \delta g\circ g^{-1}$ defines a vector field on $P$, since $\xi(p) = \delta g(g^{-1}(p))\in T_{g(g^{-1}(p))}P = T_pP$. 

Let $\theta$ be a $1$-form on $P$. Given a volume form $\rho$ on $P$, we may define a $1$-form $\alpha$ on the diffeomorphism group using the formula
\begin{align}
    \iota_{\delta g}\alpha_g = \int \iota_\xi\theta\,\rho.
\end{align}
Here $\xi = \delta g\circ g^{-1}$, as in the previous paragraph. We would like to compute the exterior derivative of $\alpha$. In order to use the formula \eqref{d_one_form}, we start by introducing a functional $I$ defined on the space of curves $g(t)$ in the diffeomorphism group with $g(-1)$ and $g(1)$ fixed. The value of $I$ at $g(t)$ is $I(g) = \int_{-1}^{1}\int \iota_{V(t)}\theta\,\rho\,dt$, where we have introduced the notation $V(t)  = (\partial_t g(t))\circ (g(t))^{-1}$. The first variation of $I$ is given by
\begin{align*}
    \delta I(g)[\delta g] &= \int_{-1}^{1}\int \iota_{\partial_t\xi+\mathcal{L}_{V}\xi}\theta\,\rho\,dt\\
    & = \int_{-1}^{1}\int \iota_{[V,\xi]}\theta\,\rho\,dt,
\end{align*}
where we have used $\iota_{\partial_t\xi}\theta\,\rho = \partial_t(\iota_{\xi}\theta\,\rho)$ and $\xi(-1) = \xi(1) = 0$.
Next we introduce 
\begin{itemize}
    \item $\delta g_1,\delta g_2$: tangent vectors to the diffeomorphism group at $g$
    \item $g(t)$: a curve of diffeomorphisms with fixed end points and $g^\prime(0) = \delta g_2$
    \item $\delta g(t)$: a vector field along the curve $g(t)$ with $\delta g(0) = \delta g_1$, $\delta g(-1) = 0$, and $\delta g(1) = 0$,
\end{itemize}
and apply the formula  \eqref{d_one_form} to obtain
\begin{align}
    \iota_{\delta g_2}\iota_{\delta g_1}d\alpha_g & = \lim_{a\rightarrow 0}\frac{1}{2a} \delta I(g)[\mathbb{I}_{[-a,a]}\delta g]\nonumber\\
    & = \lim_{a\rightarrow 0}\frac{1}{2a}\int_{-a}^{a}\int \iota_{[V(t),\xi(t)]}\theta\,\rho\,dt\nonumber\\
    & = \int \iota_{[V(0),\xi(0)]}\theta\,\rho\nonumber\\
    & = -\int \iota_{[\xi_1,\xi_2]}\theta\,\rho.
\end{align}
It is instructive to compare this calculation with the calculation of $d\Theta$ in Section \ref{Mathematical preliminaries}. Note in particular that the two calculations do not, and should not give the same result.

\section{Helmholtz-Hodge decomposition}\label{hhdecompose}

This appendix contains a reference discussion of transverse, longitudinal, and harmonic subspaces. We will present the picture for both forms and vector fields in parallel.
The fundamental theorem of Hodge theory states that the space of $k$-forms $\Omega^k$ on any closed (compact, without boundary) Riemannian manifold $M$ is equal to the $L^2$-orthogonal direct sum $\Omega^k = d\Omega^{k-1}\oplus d^*\Omega^{k+1} \oplus \Omega^k_H $, where
\begin{align}
\Omega_H = \{\alpha\in \Omega^k\mid d\alpha = 0, d^*\alpha = 0\}.
\end{align}
 We say $d\Omega^{k-1}$ is the space of \emph{exact} $k$-forms, $d^*\Omega^{k+1}$ is the space of \emph{coexact} $k$-forms, and $\Omega_H^k$ is the space of harmonic $k$-forms. The operator $d^*$ is the formal adjoint of $d$ relative to the $L^2$ inner product of forms. 

General formulas for the orthogonal projection operators into the terms of the sum $d\Omega^{k-1}\oplus d^*\Omega^{k+1} \oplus \Omega^k_H $ may be derived as follows. Suppose we have a $k$-form $\alpha = \alpha_E + \alpha_C + \alpha_H$, where $E,C,H$ denote the exact, coexact, and harmonic parts of $\alpha$, respectively. To find the exact component, we first note that the codifferential of $\alpha$ is given by $d^*\alpha  = d^*\alpha_E$. Next we note that $\alpha_E = d\lambda_C$, where $\lambda_C$ is some coexact $k-1$ form. We may therefore infer

\begin{align}
d^*\alpha = d^*d\lambda_C = (d^*d + dd^*)\lambda_C = \Delta\lambda_C,\label{solve_for_exact}
\end{align}
where $\Delta$ is the Laplace-De Rham operator. When restricted to $d\Omega^{k-1}\oplus d^*\Omega^{k+1}$, $\Delta$ has a well-defined inverse. We may therefore solve Eq.\,\eqref{solve_for_exact} uniquely for $\lambda_C$ according to $\lambda_C = \Delta^{-1}d^*\alpha$. This argument, together with an analogous argument for the coexact part of $\alpha$, shows that the exact and coexact projections, $\Pi_E,\Pi_C$ are given by 
\begin{align}
\Pi_E\alpha &= d\Delta^{-1}d^*\alpha=\Delta^{-1}dd^*\alpha\\
\Pi_C\alpha & = d^*\Delta^{-1}d\alpha=\Delta^{-1}d^*d\alpha. 
\end{align}

Now let's translate this in terms of vector calculus notation assuming $\text{dim}\,M = 3$. I will treat the cases $k=1$ and $k=2$ separately.

Every form $\alpha$ in $\Omega^1$ may be identified with a unique vector field $\bm{u}$ using the invertible mapping from vector fields to $1$-forms $\bm{u}\mapsto  \bm{u}\cdot d\bm{r}$. When either the differential or the codifferential is applied to a $1$-form, the result is 
\begin{align}
d(\bm{u}\cdot d\bm{r})&=\iota_{\nabla\times\bm{u}}d^3\bm{r}\\
d^*(\bm{u}\cdot d\bm{r})&=-\nabla\cdot\bm{u}.
\end{align}
The Laplace-De Rham operator on $1$-forms may therefore be written
\begin{align}
\Delta (\bm{u}\cdot d\bm{r}) & = -d\nabla\cdot\bm{u} + d^*\iota_{\nabla\times\bm{u}}d^3\bm{r}\nonumber\\
 & = -\nabla(\nabla\cdot\bm{u})\cdot d\bm{r} + (\nabla\times(\nabla\times\bm{u}))\cdot d\bm{r}\nonumber\\
 & = -(\nabla^2\bm{u})\,\cdot d\bm{r}.
\end{align}

Every form $\beta$ in $\Omega^2$ may be identified with a unique vector field $\bm{u}$ using the invertible mapping from vector fields to $2$-forms $\bm{u}\mapsto  \iota_{\bm{u}}d^3\bm{r}$. When either the differential or the codifferential is applied to a $2$-form, the result is 
\begin{align}
d(\iota_{\bm{u}}d^3\bm{r})&=\nabla\cdot\bm{u}\,d^3\bm{r}\\
d^*(\iota_{\bm{u}}d^3\bm{r})&=\nabla\times\bm{u}\cdot d\bm{r}.
\end{align}
The Laplace-De Rham operator on $2$-forms may therefore be written
\begin{align}
\Delta (\iota_{\bm{u}}d^3\bm{r}) & = d(\nabla\times\bm{u}\cdot d\bm{r}) + d^*(\nabla\cdot\bm{u} d^3\bm{r})\nonumber\\
 & = \iota_{\nabla\times(\nabla\times\bm{u})}d^3\bm{r} - \iota_{\nabla(\nabla\cdot\bm{u})}d^3\bm{r}\nonumber\\
 & = -\iota_{\nabla^2\bm{u}}d^3\bm{r}.
\end{align}

According to the fundamental theorem of Hodge theory, the preceding remarks imply that the space of vector fields admits two decompositions, one induced by the Hodge decomposition for $1$-forms, and the other induced by the Hodge decomposition for $2$-forms. The following argument shows that these two decompositions are essentially the same.

First consider the Hodge decomposition for $1$-forms $\Omega^1=d\Omega^0\oplus d^*\Omega^2 \oplus \Omega_H^1$. The mapping $\bm{u}\mapsto \bm{u}\cdot d\bm{x}$ is a linear isometry between $\Omega^1$ and the space of vector fields $\mathfrak{X}$. Therefore there is a corresponding $L^2$-orthogonal decomposition $\mathfrak{X} = \mathfrak{X}_E^1\oplus \mathfrak{X}_C^1 \oplus \mathfrak{X}_H^1$. The space $\mathfrak{X}_E^1$ contains all vector fields of the form $\bm{u} = \nabla \phi$, where $\phi$ is a scalar field. The space $\mathfrak{X}_C^1$ contains all vector fields of the form $\bm{u} = \nabla\times\bm{A}$, where $\bm{A}$ is a vector field. The space $\mathfrak{X}^1_H$ contains all vector fields with vanishing divergence and curl. The orthogonal projections onto these spaces are given by 
\begin{align}
\Pi^1_E\bm{u} = &\nabla [\nabla^2]^{-1}\nabla\cdot \bm{u}=[\nabla^2]^{-1}\nabla(\nabla\cdot\bm{u})\\
\Pi^1_C\bm{u} = & -\nabla\times([\nabla^2]^{-1}\nabla\times\bm{u})=-[\nabla^2]^{-1}\nabla\times(\nabla\times\bm{u})
\end{align}

 Now consider the Hodge decomposition for $2$-forms  $\Omega^2=d\Omega^1\oplus d^*\Omega^3 \oplus \Omega_H^2$. The mapping $\bm{u}\mapsto \iota_{\bm{u}}d^3\bm{x}$ is a linear isometry between $\Omega^2$ and the space of vector fields $\mathfrak{X}$. Therefore there is a corresponding $L^2$-orthogonal decomposition $\mathfrak{X} = \mathfrak{X}_E^2\oplus \mathfrak{X}_C^2 \oplus \mathfrak{X}_H^2$. The space $\mathfrak{X}_E^2$ contains all vector fields of the form $\bm{u} = \nabla\times\bm{A}$, where $\bm{A}$ is a vector field. The space $\mathfrak{X}_C^2$ contains all vector fields of the form $\bm{u} = -\nabla\phi$, where $\phi$ is a scalar field. The space $\mathfrak{X}_H^2$ contains all vector fields with vanishing divergence and curl. The orthogonal projections onto these spaces are given by
\begin{align}
\Pi^2_E\bm{u} = &-\nabla\times([\nabla^2]^{-1}\nabla\times\bm{u})\\
\Pi^2_C\bm{u} = & \nabla [\nabla^2]^{-1}\nabla\cdot \bm{u}
\end{align}
Note that we have
\begin{align}
\mathfrak{X}_E^1&=\mathfrak{X}_C^2,\quad \mathfrak{X}_C^1=\mathfrak{X}_E^2,
\end{align}
and
\begin{align}
\Pi_E^1&=\Pi_C^2,\quad \Pi_C^1=\Pi_E^2.
\end{align}
Therefore the only difference between the orthogonal decompositions induced by the Hodge decomposition for $1$-forms and the Hodge decomposition for $2$-forms is a naming convention. According to the $1$-form decomposition, the gradients are exact while the curls are coexact. According to the $2$-form decomposition, the curls are exact while the gradients are coexact. (This is actually an overstatement. When properly formulated in terms of Sobolev spaces, the two Hodge decompositions actually give slightly different decompositions for vector fields. The $1$-form decomposition expresses a vector field as the sum of a strong gradient, a weak curl, and a Harmonic field, while the $2$-form decomposition expresses a vector field as the sum of a strong curl, a weak gradient, and a Harmonic field. This distinction will not play an important role in this work.)  

In light of the previous remarks, it is convenient to introduce a separate notation for the decomposition of vector fields. We write $\mathfrak{X}_L = \mathfrak{X}_E^1 = \mathfrak{X}_C^2$ for the transverse subspace and $\mathfrak{X}_T = \mathfrak{X}_C^1 = \mathfrak{X}_E^2$ for the longitudinal subspace. The transverse and longtiudinal projections are then
\begin{align}\label{PiLEquals}
\Pi_L\bm{u} = &\nabla [\nabla^2]^{-1}\nabla\cdot \bm{u}=[\nabla^2]^{-1}\nabla(\nabla\cdot\bm{u})\\
\Pi_T\bm{u} = & -\nabla\times([\nabla^2]^{-1}\nabla\times\bm{u})=-[\nabla^2]^{-1}\nabla\times(\nabla\times\bm{u}).\label{PiTEquals}
\end{align}


In summary, every vector field on $M = T^3$ has the unique decomposition $\mathbf{u} = \mathbf{u}_T  + \mathbf{u}_L + \mathbf{u}_H$, where
\begin{equation}
    \mathbf{u}_T = \Pi_T\mathbf{u} = -\nabla\times([\nabla_{EC}^2]^{-1}\nabla\times\mathbf{u})
\end{equation}
and
\begin{equation}
    \mathbf{u}_L = \Pi_L\mathbf{u} = \nabla([\nabla_{EC}^2]^{-1}\nabla\cdot\mathbf{u})
\end{equation}
and 
\begin{equation}
\mathbf{u}_H = \Pi_H\mathbf{u} = \int\mathbf{u}\,d^3\mathbf{r}/ \int\,d^3\mathbf{r}.
\end{equation}
Here $\nabla^2_{EC}$ is either the vector Laplacian restricted to vectors of the form $\nabla\phi +\nabla\times\mathbf{A}$ or the scalar Laplacian restricted to scalars of the form $\nabla\cdot\mathbf{A}$.

\section{All-orders existence of rectifying transformation\label{rect_exist_sec}}
This Appendix proves that the rectifying transformation discussed in Section \ref{Derivation of the rectifying transformation} exists to all orders in $\epsilon= 1/c$. In this Appendix only, if $\omega$ is a $2$-form then $\widehat{\omega}$ denotes the bundle map $TM\rightarrow T^*M:v\mapsto \iota_v\omega=: \widehat{\omega}\,v$. Similarly, if $j$ is a bivector then $\widehat{j}$ denotes the bundle map $T^*M\rightarrow TM:\alpha\mapsto \iota_{\alpha}j=: \widehat{j}\,\alpha$. Note that if $\omega$ is a symplectic form with corresponding Poisson bivector $j$ then $\widehat{\omega}^{-1} = -\widehat{j}$.

Let $\omega_\epsilon = \omega_0 + \epsilon\,\omega_1 + \dots$ be a formal power series in $\epsilon$ whose coefficients are exact $2$-forms $\omega_k = -d\theta_k$. Assume there exists a formal power series $j_\epsilon = j_0 + \epsilon\,j_1 + \dots$ with bivector coefficients such that $-\epsilon^{-1}\widehat{j}_\epsilon\,\widehat{\omega}_\epsilon =  \text{id}_{TM}$ in the sense of formal power series. Here $\text{id}_{TM}$ denotes the identity map $TM\rightarrow TM$. We remark that the formal power series $2$-form whose first coefficients are given in Eqs.\,\eqref{dtheta_0}-\eqref{dtheta_3} satisfies this assumption. We leave it as an exercise for the reader to verify this claim. We would like to show that there exists a sequence of Lie transforms with generating vector fields $G_k$, $k\geq 1$, such that $\dots\exp(-\mathcal{L}_{G_2})\exp(-\mathcal{L}_{G_1})\omega_\epsilon = \omega_0 + \epsilon\,\omega_1$. Such a sequence of Lie transforms comprises an all-orders rectifying transformation.

As a first step, we will show that there exists a unique formal power series vector field $G_1$ that solves the equation
\begin{align}
    \iota_{G_1}(\omega_0 + \epsilon\,\omega_1) = -\epsilon^2\theta_2.\label{G1_eqn_gen}
\end{align}
Let $\epsilon^2\delta\omega_\epsilon = \omega_\epsilon - (\omega_0 + \epsilon\,\omega_1) = \epsilon^2(\omega_2 + \epsilon\,\omega_3 + \dots)$. Applying $-\epsilon^{-1}\widehat{j}_\epsilon$ to the identity $\widehat{\omega}_\epsilon = (\widehat{\omega}_0 + \epsilon\,\widehat{\omega}_1) + \epsilon^2\,\delta\widehat{\omega}_\epsilon$ implies $\text{id}_{TM} = -\epsilon^{-1}\widehat{j}_\epsilon (\widehat{\omega}_0 + \epsilon\,\widehat{\omega}_1) - \epsilon\,\widehat{j}_\epsilon \delta\widehat{\omega}_\epsilon$, which is equivalent to 
\begin{align}
    \text{id}_{TM} = -\epsilon^{-1}(\text{id}_{TM} + \epsilon\,\widehat{j}_\epsilon \delta\widehat{\omega}_\epsilon)^{-1}\widehat{j}_\epsilon (\widehat{\omega}_0 + \epsilon\,\widehat{\omega}_1),\label{formal_invertibility}
\end{align}
where $(\text{id}_{TM} + \epsilon\,\widehat{j}_\epsilon \delta\widehat{\omega}_\epsilon)^{-1} = \text{id}_{TM} - \epsilon\,\widehat{j}_0\,\delta \widehat{\omega}_0+\dots$. The formula \eqref{formal_invertibility} says $\widehat{\omega}_0 + \epsilon\,\widehat{\omega}_1$ has a formal inverse given by $-\epsilon^{-1}(\text{id}_{TM} + \epsilon\,\widehat{j}_\epsilon \delta\widehat{\omega}_\epsilon)^{-1}\widehat{j}_\epsilon$. Applying this formal inverse to both sides of \eqref{G1_eqn_gen} therefore reveals the unique $G_1$ that satisfies \eqref{G1_eqn_gen}, namely
\begin{align}
    G_1 = \epsilon\,(\text{id}_{TM} + \epsilon\,\widehat{j}_\epsilon \delta\widehat{\omega}_\epsilon)^{-1}\widehat{j}_\epsilon\,\theta_2.
\end{align}
Note that $G_1$ is an $O(\epsilon)$ formal power series in $\epsilon$. Also observe that 
\begin{align}
    \exp(-\mathcal{L}_{G_1})\omega_\epsilon &= \omega_\epsilon -d\iota_{G_1}\omega_\epsilon + \frac{1}{2}d\iota_{G_1}d\iota_{G_1}\omega_\epsilon +d[O(\epsilon^3)]\nonumber\\
    & = \omega_0 + \epsilon\,\omega_1 - \epsilon^2\,d\theta_2\nonumber\\
    & -d\iota_{G_1}(\omega_0 +\epsilon\,\omega_1) + \frac{1}{2}d\iota_{G_1}d\iota_{G_1}(\omega_0 + \epsilon\,\omega_1) + d[O(\epsilon^3)]\nonumber\\
    & = \omega_0 + \epsilon\,\omega_1 -\epsilon^2\frac{1}{2}d\iota_{G_1}d\theta_2\nonumber + d[O(\epsilon^3)]\\
    & = \omega_0 + \epsilon\,\omega_1  + d[O(\epsilon^3)].\label{initial_step}
\end{align}

Now we will prove existence of the sequence $G_k$, $k\geq 1$, by induction. Suppose that there is a sequence of $G_k$, $1\leq k \leq n$ such that $\exp(-\mathcal{L}_{G_n})\dots\exp(-\mathcal{L}_{G_1})\omega_\epsilon = \omega_0 + \epsilon\,\omega_1 + \epsilon^{2+n}\,\beta_\epsilon$, where $\beta_\epsilon^{(n)} = \beta_0^{(n)} + \epsilon\,\beta_1^{(n)} + \dots$ is a formal power series whose coefficients are exact $2$-forms $\beta_k^{(n)} = -d\alpha_k^{(n)}$. Note that the previous paragraph established existence of such a sequence with $n=1$. We would like to show that there exists a $G_{n+1}$ such that $\exp(-\mathcal{L}_{G_{n+1}})\exp(-\mathcal{L}_{G_n})\dots\exp(-\mathcal{L}_{G_1})\omega_\epsilon = \omega_0 + \epsilon\,\omega_1 + \epsilon^{2+(n+1)}\,\beta_\epsilon^{(n+1)}$, where $\beta_\epsilon^{(n+1)} = \beta_0^{(n+1)}+\epsilon\,\beta_1^{(n+1)}+\dots$ is a formal power series whose coefficients are exact $2$-forms. We define $G_{n+1}$ by requiring that is solves the linear equation
\begin{align}
    \iota_{G_{n+1}}(\omega_0 + \epsilon\,\omega_1) = - \epsilon^{2+n}\alpha_0^{(n)},
\end{align}
whose unique solution is
\begin{align}
    G_{n+1} = \epsilon^{n+1}(\text{id}_{TM} + \epsilon\,\widehat{j}_\epsilon \delta\widehat{\omega}_\epsilon)^{-1}\widehat{j}_\epsilon\,\alpha_0^{(n)}.
\end{align}
This $G_{n+1}$ has the required properties since
\begin{align}
&\exp(-\mathcal{L}_{G_{n+1}})\exp(-\mathcal{L}_{G_n})\dots\exp(-\mathcal{L}_{G_1})\omega_\epsilon   \nonumber\\
=&\exp(-\mathcal{L}_{G_{n+1}})\bigg(\omega_0 + \epsilon\,\omega_1 - \epsilon^{2+n}d\alpha_0\bigg) + d[O(\epsilon^{2+(n+1)})]\nonumber\\
=&\omega_0+\epsilon\,\omega_1 - d\iota_{G_{n+1}}(\omega_0 + \epsilon\,\omega_1) - \epsilon^{2+n}d\alpha_0+ d[O(\epsilon^{2+(n+1)})]\nonumber\\
=&\omega_0+\epsilon\,\omega_1+\epsilon^{2+n}d\alpha_0^{(n)} -\epsilon^{2+n}d\alpha_0+ d[O(\epsilon^{2+(n+1)})]\nonumber\\
=&\omega_0+\epsilon\,\omega_1+ d[O(\epsilon^{2+(n+1)})].
\end{align}

\bibliographystyle{jpp}

\bibliography{biblio}

\end{document}